# CoRTeC: A Differentially Private Release for Synthetic Tabular Data, Decodable by Any Generator

**Calister Nnona**
research@calisternnona.com

**Abstract.** The records that could train the most useful models, such as a hospital's patient files, are often the ones an institution cannot share. We propose CoRTeC, a method that releases synthetic versions of such records under differential privacy. Differential privacy places a proven limit on what the synthetic data can reveal about any person, and in particular on membership inference. CoRTeC spends the privacy budget once, on a small file of noisy counts designed for training a classifier. For each group of records, the file gives the group's size, its balance of outcomes, a histogram of every column for each outcome and the outcome rate in each of several non-overlapping cells. A reviewer can check every number before the file leaves the institution. Anyone can then turn the counts into synthetic rows, as often as they like and at no further privacy cost. A maximum-entropy decoder can do this without any language model, and so can each of the five frozen language models we tested. None of them ever sees a real record. A selection step then matches the rows to the released counts. Decoded without a language model, the automatically configured release at $\varepsilon = 2$ trains classifiers to within 0.02 AUC of a real sample of the same size on the Adult dataset, across three data splits. On NHANES (three splits) and bank marketing (one split), rerun with freshly derived configurations and the selection step, it stays within 0.04 AUC on average. On Adult, with a fixed cohort rule, the release carries the outcome rates a deployment names more closely than a real sample does, at some cost in accuracy on pairs of columns. Four controls show that most of the utility comes from the release and the selection step, and that a language model adds what it knows from the column names. No private row ever reaches a model, and no row CoRTeC generates copies a real one. The strongest membership attack stays far below the bound the budget allows.

## 1. Introduction

Hospitals, insurers and statistics offices hold sensitive tables that they would like to use more widely. They want to train models outside their own walls, give realistic samples to partners and let researchers study a problem without studying the people in the data. Differentially private [14, 15] synthetic data [6, 27] is a widely accepted way to do all three. Today, it comes from two kinds of method. Marginal methods such as MST [27] and AIM [28] fit a graphical model to marginals measured under the privacy budget [10]. Methods built on language models either fine-tune a model on the private rows with DP-SGD [1, 3, 38] or score a model's candidates privately [24, 43].

An institution with sensitive data needs three things from such a method. First, it needs to spend its budget once, on counts it chose in advance and can review before anything leaves its walls. Second, it needs to be able to decode those counts again later, with another team or another model, at no further privacy cost. Third, when a language model is involved, it needs to know what the model adds beyond the private statistics. A model pretrained on public text may already know a public benchmark [4, 10, 37]. Most existing synthetic-data methods spend the budget as they go: inside a fitted model, during a training run or over rounds of candidates. They are not designed to give these three things.

In this paper, we propose CoRTeC: a differentially private release designed for training a downstream classifier, together with two ways of decoding it (Figure 1). Apart from the optional Stage C, Stage A is the only code that reads the private data. It groups the rows into cohorts and publishes four things for each cohort, all with Laplace noise: the cohort's size, its class balance, a histogram of every attribute for each outcome class and the outcome rate in each of a set of non-overlapping cells. The workload is fixed before the release is computed, either by a public rule or by a charged column ranking whose output is itself released (Section 4.1). The budget is spent there and only there. What leaves the institution is a file of counts that anyone can inspect cell by cell. Everything after it is post-processing. A maximum-entropy decoder (the same estimator AIM uses) can fit the released counts and sample rows from them. Alternatively, a frozen language model can decode the same release. It reads only the release and writes rows to match the exact counts it is asked for. A selection step then takes the rows from either decoder, matches them to the released counts and redraws values inside their released bins. Because everything that reads only the release is post-processing, generation can be repeated as often as needed, on any model, even one behind an interface the institution does not control. The model becomes a replaceable part, and replacing it costs no privacy.

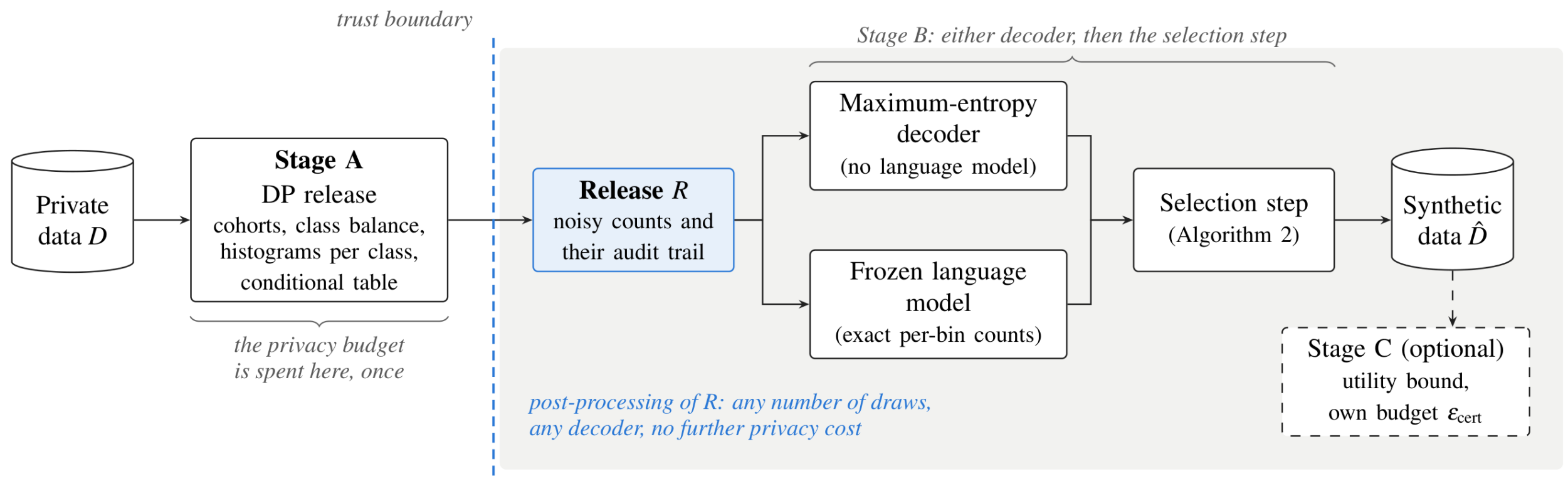

**Figure 1:** The privacy budget is spent once in Stage A. In Stage B, either the maximum-entropy decoder or a frozen language model turns the released statistics into rows, and the selection step matches those rows to the released counts. Everything to the right of the dashed line except Stage C reads only the release, so it can be repeated without limit.

**What an institution gains.** The design gives an institution five things, some of which marginal methods and Private Evolution also offer [24, 26]. First, the entire privacy loss sits in one file in which every number comes with its query, its budget and how it composes. Reviewing a release means reviewing counts, not model weights. Second, the workload is fixed before the release is computed, rather than discovered round by round. Third, only the release crosses the institution's boundary, so even a hostile endpoint learns nothing the release does not already say. Fourth, decoding is free and can be repeated at a different size, with a newer model or with no model at all. Fifth, because the model is separate from the mechanism, its contribution can be measured.

We evaluate CoRTeC on three datasets: Adult, NHANES and bank marketing. Every result is read against a real sample of the same size, a floor with the target shuffled and a ceiling trained on real data. Four controls separate the parts of the design. Two decode the release without a language model, with and without the selection step. One renames the schema to neutral tokens, and one writes rows from the column names with no release at all. As reference points, we also run AIM, the main one, and MST, as well as three recent methods under their own protocols. Every table reports fidelity and utility side by side, and the central results are repeated on three splits.

With the fixed cohort rule and the selection step, the decoded rows carry the conditional rates the release names more closely than a real sample of the same size does. The cost of fixing the workload in advance falls on pairs of columns, which an adaptive workload fits more closely. Rerun with freshly derived configurations and decoded with the maximum-entropy decoder and the selection step, the releases for NHANES and bank marketing stay within 0.04 AUC of a real sample on average over their splits. In every CoRTeC configuration, no private row reaches the model, and no synthetic row copies a training row.

Our contributions are as follows.

- **A release built around what a classifier needs** (Section 4.1). Stage A spends the whole budget once, on class-conditional histograms within cohorts formed by a public rule or a charged column ranking and on a conditional outcome-rate table, all fixed before the release is computed. The release mechanism satisfies pure ε-differential privacy with a five-step proof (Section 4.3), and a reviewer can check it cell by cell before it leaves the institution. Fixed workloads and published noisy measurements are established practice [48, 52]. What we add is a release designed for a downstream classifier and the conditional rates a deployment names. An auto-configurator chooses the cohorts and the conditional levels within the same budget.
- **A frozen language model as a decoder of that release** (Section 4.2). The model is prompted with exact per-bin, per-category and per-class counts derived from the release and never sees a record. A selection step (Algorithm 2) then keeps whole generated rows that match the released counts and redraws values inside their released bins. Swanberg et al. give a frozen model the column names alone and reweight its rows to noisy counts [37], and Ito et al. give it noisy summary statistics of educational time series [50]. Because every decoder reads only the release, the model can run behind a vendor's interface and be replaced without touching the guarantee. A maximum-entropy decoder [26] decodes the same release with no language model.
- **Utility without fine-tuning** (Section 5). Decoded without fine-tuning any model, the release trains classifiers close to a real sample of the same size: within 0.02 AUC on Adult across three splits with the maximum-entropy decoder. Under the protocols of three recent DP generators, its utility is within 1.4 percentage points of their published results or above them. The API-only methods of Swanberg et al. did not improve on the marginal baselines [37], and the authors of DP-LLMTGen conclude that "fine-tuning is necessary for tabular data" [38].

- **Evidence of what each part contributes** (Section 5). Controls show that most of the utility comes from the release and the selection step, and that a language model adds what it knows from the column names.
- **A reference implementation and a deployment guide.** The implementation ships both decoders (`cortec run` and `cortec decode`), an optional bound on the conditional rates (Stage C) in the spirit of AIM's error bounds [28] and the checks a deployment needs: membership inference tested on a positive control, distance to the closest record, a memorisation test and the cost per thousand rows. The deployment guide gives the reference architecture, the guardrails and the failure modes we met, each with the test that catches it.

## 2. Related Work

Three families of methods address this task. Some measure marginals under the budget, some train a generator under the budget and some keep a foundation model frozen. In Section 5, we run the members of these families that an institution would be most likely to consider: the reference AIM and MST, DP-2Stage, the DP-LLMTGen procedure on three backbones and Tab-PE.

**Marginal methods.** PrivBayes [46], MST [27], PrivMRF [7] and AIM [28] select low-order marginals under the budget and fit a graphical model to them. MST and AIM use Private-PGM [26] for that fit. NIST-MST, the predecessor of MST, took first place in the 2018 NIST challenge [27]. Bowen and Snoke compare that challenge's entries on marginal, joint and correlation measures [6]. AIM chooses its marginals adaptively and is our main reference from this family. The benchmark of Chen et al. [10] finds AIM and PrivMRF the most accurate across five datasets and three budgets, measured by machine-learning efficacy, 3-way query error and 2-way total variation. In a 2021 benchmark, Tao et al. [36] found the marginal-based methods ranked highest, with MST first. Fixed workloads answered once are long established, from HDMM [52] to the published noisy measurements of the 2020 US Census [48]. PrivSyn adjusts synthetic records until they match noisy marginals [54], and AIM bounds the error of its synthetic marginals from its own measurements at no extra cost [28]. Our AIM and MST rows use the authors' private-pgm implementation on our public bins.

**Generators trained under differential privacy.** DP-GAN [44] and DP-CTGAN [32], which builds on CTGAN [42], train a generative adversarial network with DP-SGD [1], and PATE-GAN [22] trains one under PATE [30, 31]. DP-TBART [8] trains an autoregressive transformer with DP-SGD, and DP-TLDM [47] trains an autoencoder with DP-SGD and a diffusion model on its latents. DP-2Stage and DP-TBART evaluate several of them beside marginal methods [3, 8], and Chen et al. leave the adversarial generators out of their benchmark for their limited performance in other comparisons [10]. We therefore cite them rather than run them.

**Language models fine-tuned on the private rows.** GReaT [5] writes each row as text and fine-tunes GPT-2 on it without a privacy guarantee. DP-LLMTGen [38] fine-tunes Llama-2 7B in two stages: first on random tables to learn the format, then on the private rows with DP-SGD and a loss weighted towards the value tokens. Its authors find that meaningful feature names help. DP-2Stage [3] also fine-tunes GPT-2 in two stages, the first on pseudo-data, and it includes AIM among its reference points on Adult. In DP-LLMTGen and DP-2Stage, the private rows reach the training hardware, and the budget is spent on per-example gradients in every epoch.

**Methods that keep the model frozen.** Private Evolution [24] and its text version [43] generate candidates with a foundation model, let the private records vote for their nearest candidates under a Gaussian mechanism and repeat. Tab-PE [39] replaces the model with simple variation operators. It also shows that marginal methods miss high-order correlations that the standard benchmarks do not test. Swanberg et al. [37] adapt Private Evolution to tabular data. They also generate rows in one shot from the column names and use them as public data, either inside a marginal mechanism or reweighted to match a fixed DP workload. Their Private Evolution variant, run on New York taxi data, does not improve on simple baselines. On Adult, the one-shot rows do not improve on MST and are generally inaccurate on pairs of columns [37]. Closest to CoRTeC, Ito et al. prompt frozen models with Gaussian-noised summary statistics of educational time series and treat generation as post-processing [50]. DP-TabICL places noisy per-group aggregates in the prompt of a frozen model, for in-context classification rather than generation [49]. We run Tab-PE [39] and the one-shot setting as well. CoRTeC differs from these in what the model is given and what follows: a release of class-conditional statistics with exact counts, which the selection step then enforces. Table 1 summarises the differences.

**Table 1:** What each method needs. The budget column says where the privacy loss is incurred.

| method | model weights | private rows seen by | public or side data | budget spent on | adaptive rounds | backbone swap |
|---|---|---|---|---|---|---|
| CoRTeC | none (any inference interface) | Stage A only (the model never sees a row) | column names and value lists | one release of counts, composed once | none | at no privacy cost |
| AIM, MST | none | the measured marginals | public domains and bins | measured marginals (AIM selects adaptively) | AIM yes | not applicable |
| DP-2Stage | GPT-2, full fine-tune | the DP-SGD trainer | pseudo data (uniform over private ranges, or a public table) | per-sample gradients, every step | yes | retrain |
| DP-LLMTGen | Llama-2 7B, LoRA | the DP-SGD trainer | random tables from names, categories and ranges | per-sample gradients, every step | yes | retrain |
| Tab-PE | none | the nearest-neighbour histogram, every iteration | numerical bounds and class counts | Gaussian noise on the histogram, 15 to 30 rounds | yes | not applicable |
| one-shot records [37] | none | the downstream mechanism, not the model | column names and types | inside the downstream mechanism | depends | at no privacy cost |

## 3. Preliminaries and Privacy Model

A dataset $D$ is a multiset of $n$ records over attributes $A = A_{num} \cup A_{cat}$ with a binary target $y$. Each attribute has a public domain: an interval for a numerical attribute and a finite set of values for a categorical one. Public bin edges divide each numerical domain into bins. $M_a(D)$ denotes the histogram of attribute $a$ over its bins. For a cell $c$ of the record space, $k_c(D)$ is the number of records in $c$ with $y = 1$. Two datasets are neighbours if one can be obtained from the other by adding or removing a single record.

**Definition 1 (Differential privacy [14]).** *A randomised mechanism $M$ is ε-differentially private if for all neighbours $D$, $D'$ and all sets $S$ of outputs, $Pr[M(D) \in S] \leq e^{\varepsilon} Pr[M(D') \in S]$.*

The Laplace mechanism adds Lap(Δ/ε) noise to a function of L1 sensitivity Δ and is ε-differentially private [14]. A histogram over any number of bins has $\Delta = 1$, and so does a counting query. Mechanisms run on the same data compose sequentially: their budgets add up [15]. Mechanisms run on disjoint parts of the data compose in parallel, provided that the parts are defined by a rule that does not depend on the data. Together, they cost only the largest of their budgets [29]. Any computation on the output of a private mechanism that does not look at $D$ again is post-processing and costs nothing [15]. Definition 1 also limits what any membership attacker can achieve. No test of whether a record is in $D$ can have a true-positive rate that exceeds its false-positive rate by more than $(e^{\varepsilon} - 1)/(e^{\varepsilon} + 1)$, which is 0.762 at $\varepsilon = 2$.

**Threat model.** The adversary has unlimited computing power and sees everything that crosses the trust boundary: the release $R$, every synthetic dataset drawn from it and the model's inputs and outputs. The institution's data extract, the schema and the release engine all stay inside its perimeter, and the release engine, with the optional Stage C, is the only code that reads $D$. The generator, the selection step and everything downstream read only $R$ and their own output. Whoever operates the model therefore learns nothing beyond what $R$ reveals, even if they are hostile, and $R$ itself is ε-differentially private.

**Privacy model.** The guarantee is pure ε-differential privacy ($\delta = 0$) in the central model under add/remove adjacency. It protects one row. When each person contributes one row, it protects a person. When a person can contribute several rows, as with hospital encounters, the per-person ε is multiplied by the largest number of rows any one person contributes. The reference implementation therefore requires that number to be declared. Two things are treated as public: the list of attributes with their domains and bins, as in every marginal method, and the rule that forms the cohorts (Section 6 notes where our runs read the category levels from the data). Cohort sizes, class balances and cell sizes are private counts, so they are released with noise rather than assumed. The model's pretraining corpus falls outside the guarantee. If a private record was in it, that record was exposed before CoRTeC ever ran. Like Swanberg et al. [37], we treat a publicly released pretrained model as public information, an assumption Tramèr et al. question [40]. We measure the consequence that matters: whether anything the model has memorised reaches a draw (Section 5.5).

## 4. CoRTeC: the Release and its Decoders

CoRTeC has three stages (Figure 1). Apart from the optional Stage C, Stage A is the only code that reads $D$. It produces the release $R$. Stage B decodes $R$ with either the maximum-entropy decoder or a frozen language model and passes the decoder's output through a selection step. It reads only $R$ and its own output. Stage C is optional and has its own budget. It bounds how far the synthetic conditional rates are from the private ones.

### 4.1 Stage A: the release

**Algorithm 1: CoRTeC-Release**

**Input:** private $D$; public schema (domains, bin edges, stratification rule $\Pi$, conditional levels $L_0 \subset \dots \subset L_k$); budget $\varepsilon_{total}$; conditional share $\alpha$; count share $\gamma$; suppression floor $n_{min}$

**Output:** release $R$

(1) $\varepsilon_{count} \leftarrow \gamma\, \varepsilon_{total}$; $\varepsilon_{marg} \leftarrow (1 - \alpha)(\varepsilon_{total} - \varepsilon_{count})$; $\varepsilon_{cond} \leftarrow \alpha\, (\varepsilon_{total} - \varepsilon_{count})$

(2) $q \leftarrow |A| + 1$ queries per cohort; $\varepsilon_q \leftarrow \varepsilon_{marg} / q$; $\varepsilon_L \leftarrow \varepsilon_{cond} / |L|$

(3) For each cohort $P$ the rule $\Pi$ can form, holding records or not:

  (4) $\tilde{n}_P \leftarrow |P| + \mathrm{Lap}(1/\varepsilon_{count})$; if $\tilde{n}_P < n_{min}$ continue; release $\tilde{n}_P$

  (5) release the two-bin histogram of $y$ on $P$, each count + $\mathrm{Lap}(1/\varepsilon_q)$

  (6) if both noised class counts are at least $n_{min}$: for $c \in \{0, 1\}$ and $a \in A$ release $M_a(P \cap \{y = c\}) + \mathrm{Lap}(1/\varepsilon_q)$

  (7) else: for $a \in A$ release $M_a(P) + \mathrm{Lap}(1/\varepsilon_q)$ (pooled block)

(8) For each level $\ell \in L$ and each cell $c$ of its public domain, holding records or not:

  (9) $\tilde{k}_c \leftarrow k_c(D) + \mathrm{Lap}(2/\varepsilon_L)$; $\tilde{n}_c \leftarrow |c| + \mathrm{Lap}(2/\varepsilon_L)$; if $\tilde{n}_c < n_{min}$ continue

  (10) release $\tilde{\rho}_c \leftarrow \mathrm{clip}(\tilde{k}_c / \max(\tilde{n}_c, n_{min}), 0, 1)$

(11) Return $R$ with every released quantity, its $\varepsilon$, sensitivity and composition rule

Four design choices keep the release cheap. First, the cohorts come from a rule over public domains, so forming them costs nothing. Because the cohorts do not overlap, their queries also compose in parallel. Second, the release uses histograms rather than means and standard deviations. A histogram has sensitivity 1 however many bins it has, so the whole shape of a distribution costs the same as a single count. Third, giving each outcome class its own block of histograms costs no more than a single pooled block, because the two class blocks do not overlap. The pooled histogram is simply their mixture under the released class balance, which is post-processing. A cohort whose smaller class falls below $n_{min}$ keeps a pooled block instead. Fourth, the conditional table is released over the cells of a partition, so each level receives its full budget $\varepsilon_L$ however many cells it contains. Each cell's rate is the ratio of two noised counts whose noise does not depend on anything private. This keeps the mechanism pure $\varepsilon$-differentially private. Noising the rate directly would need noise scaled to the private cell size, which would break pure $\varepsilon$-differential privacy. Every decision involving $n_{min}$ compares a count that has been noised and charged to the budget with the public floor. Which cohorts and cells appear in $R$ therefore depends only on released quantities. An auto-configurator can also choose $\Pi$ and $L$ from the schema, the budget and the requested size. Its only step that reads $D$ is a noised contingency table for each candidate column, and that step is charged to the budget (Appendix D).

**Example 1 (one cohort of a release).** Take the seed-42 Adult release at $\varepsilon = 2$. Every fixed-rule Adult configuration in Section 5 decodes this release, the anonymised ones under neutral names. The release forms twelve cohorts from four education bands and three bands of weekly hours. Consider the cohort of graduates who work forty hours a week. With $\gamma = 0.02$, the cohort-size query has $\varepsilon_{count} = 0.04$, so Stage A releases the true size plus Laplace noise at scale 25. The released size is 762. The class-balance query and the fourteen histograms share $\varepsilon_{marg} = 1.568$. That makes fifteen queries at $\varepsilon_q = 0.105$ each, so every count carries noise at scale 9.6. The two noised class counts give a balance of 0.551 above 50K and 0.449 at or below it. Both counts exceed $n_{min} = 150$, so the cohort receives a separate block of histograms for each class. In the block above 50K, age puts 0.216 of its mass in the 40 to 45 bin and none below 25. Relationship is 0.744 husband and 0.179 not in a family. In the block at or below 50K, age puts 0.140 in the 25 to 30 bin and 0.049 below 25, while relationship is 0.282 husband and 0.419 not in a family. A category whose noised share is 0.005 or less is listed as floored and released at a fixed floor share instead, so that it can still be generated. The floored categories of a column share 1 % of its mass equally. This is post-processing of the noised histogram. Here, that applies to three categories above 50K (other relative, own child and unmarried), each released at 0.0033, and to other relative at or below it, released at 0.01. The two released levels, positions 0 and 3 of the declared hierarchy, share $\varepsilon_{cond} = 0.392$. Level 0 is a single cell holding the overall rate of 0.241. Level 3 crosses education, hours and marital status, which gives this cohort two cells: 0.739 for married people and 0.210 for everyone else. Each rate is the ratio of two counts noised at scale 10.2, with the denominator held at or above the public floor. That is everything about this cohort that leaves the institution: a size, two class counts, twenty-eight histograms, a list of floored categories and two rates, each with its ε recorded beside it.

Figure 2 shows the two class blocks of this cohort as released, and Appendix D follows the cohort through Stage B.

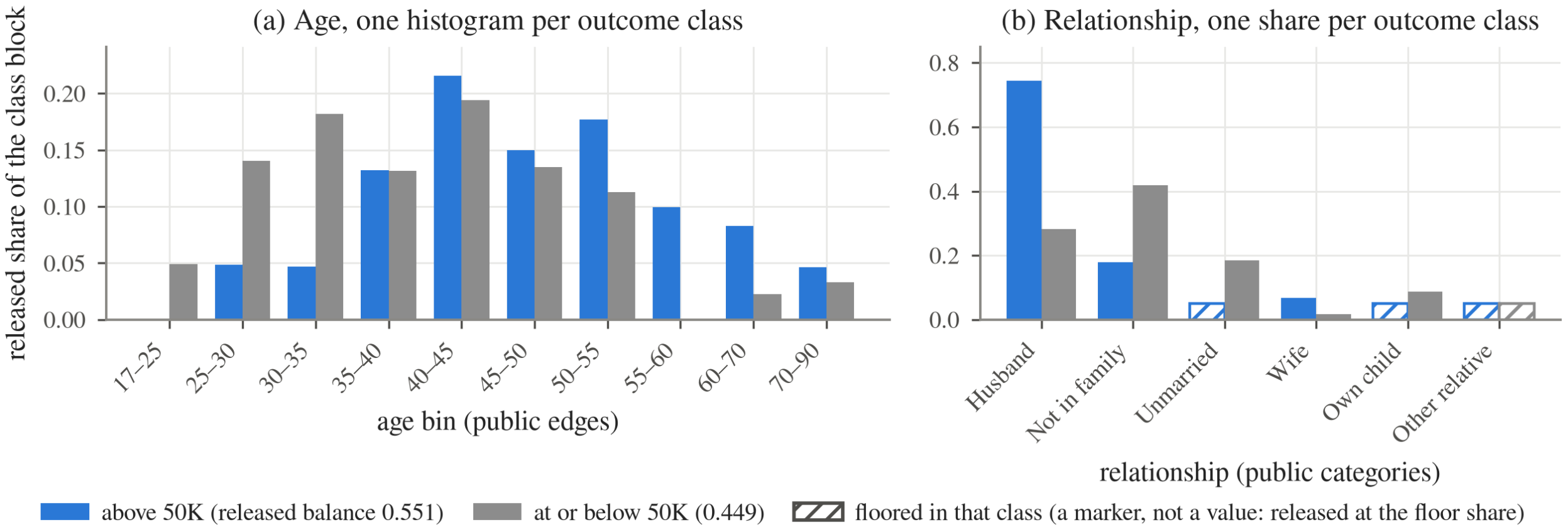

**Figure 2:** The cohort of Example 1 as it leaves the institution. (a) The age histogram released for each outcome class over the public bin edges. (b) The relationship shares for each class. Hatched markers show categories whose noised share fell to 0.005 or below. These are released at a fixed floor share (Example 1). Every bar is a released number. The decoders read only released numbers like these, together with the class balance, the cohort size and the two conditional cells.

### 4.2 Stage B: exact counts, selection and the sub-bin rule

Stage B divides the requested rows among the cohorts in proportion to their released sizes. From the released histograms, it works out exact counts for every bin, category and class. When the decoder is a language model, Stage B then asks the frozen model for batches of 25 rows that carry whatever counts the cohort still owes. The prompt (Appendix D) contains the release and the counts but never a record. It states the counts separately for each outcome class. It also tells the model to follow the released statistics even where they contradict its expectations. Rows that parse, lie within their domains and use declared categories go into a pool $k$ times the requested size, with $k = 3$ on Adult and bank marketing and $k = 2$ on NHANES. Generation runs cohort by cohort rather than cell by cell. A cell-by-cell variant we tried produced nothing for the people the released cells did not name. It dropped whole subgroups while every aggregate check still passed. This failure resembles the disparate impact that Ganev et al. document for differentially private synthesis [17], although here the cause is the generation design rather than the noise. The reference implementation refuses that path below a coverage floor (see the deployment guide).

---

**Algorithm 2: CoRTeC-Select**

---

**Input:** release $R$; pool $\hat{D}$; requested size $m$

**Output:** synthetic dataset $D^*$ of $m$ rows

(1) Require that every cohort $P$ holds at least $m\,\tilde{n}_P / \Sigma\,\tilde{n}$ pool rows and a tenth more (coverage guard)

(2) $w \leftarrow$ inclusion weights by iterative proportional fitting [12] of $\hat{D}$ to every released cell that expects at least one row: cohort sizes, class balances and every histogram, with floored categories at their floor share; a row carrying a category absent from the release receives weight 0

(3) $D^* \leftarrow$ a systematic sample of $m$ rows under $w$ (rake)

(4) Swap a row of $D^*$ for a pool row whenever the swap lowers the weighted L1 distance between the cell counts of $D^*$ and $R$, until no swap improves it (refine)

(5) In every cohort with class blocks, redraw each numerical value uniformly inside its released bin intersected with its stratification band; integer columns draw integers (sub-bin rule)

(6) Return $D^*$

---

Exact counts make the pool match the release closely. Even so, any $m$ rows drawn from a faithful pool still carry sampling error. The selection step largely removes it. The sub-bin rule then handles the structure inside each bin. A released histogram fixes the count in each bin and nothing finer. Where a value falls within its bin is therefore not for the model to decide, so the rule redraws it uniformly. Each step reads only $R$, the generated rows and fresh randomness. Whether a numerical column holds integers is decided from the public bin edges and bounds together with the majority of the pool's values. A rule that looked at every value let a single model-written "45.0" turn a whole column into real-valued redraws. Every CoRTeC draw in this paper is scored after its integer columns are floored (Appendix D).

**The maximum-entropy decoder.** CoRTeC's second decoder needs no language model at all. It uses Private-PGM [26] to fit the graphical model whose marginals best match the released counts. That model is the maximum-entropy distribution with those marginals, and the decoder samples rows from it. Since it reads only $R$, it spends no budget. Section 5 also uses it without the selection step to measure what a language model adds. Private-PGM is also the estimator behind AIM, so on this path what CoRTeC contributes is the workload, not the inference engine. The release is a fixed workload of class-conditional single-column marginals within cohorts plus the conditional cells, all chosen before the release is computed. It gives a deployment the conditional rates it names and a file it can inspect cell by cell, and it is computed in seconds. An adaptive workload fits pairs of columns more closely (Section 5.4).

**Stage C.** For each released cell, the private data has a positive rate $p_c$, and the synthetic data has a public rate $q_c$. Stage C releases a Laplace estimate of $p_c$ at a declared budget $\varepsilon_{\mathrm{cert}}$, with the denominator held at or above the public floor. It turns that estimate into a bound on $|p_c - q_c|$ that holds for all cells at once. It gives no verdict unless two conditions are met. First, a real hold-out sample must pass the tolerance while the same sample with its target shuffled fails it. Second, every cell must hold at least 20 synthetic rows. Stage C bounds the disjoint cells of one released level. Its verdict also reads a real hold-out sample without noise, so the verdict sits outside the guarantee and should stay inside the institution. The bound concerns utility, not privacy (Appendix E.8). AIM bounds the error of its synthetic marginals in a similar way [28]. Stage C applies the idea to the conditional rates, with a fresh estimate at its own budget and the check above.

### 4.3 Privacy analysis

**Theorem 1.** *Algorithms 1 and 2 together with the generation step are $\varepsilon_{total}$-differentially private. With auto-configuration, the total is $\varepsilon_{sel} + \varepsilon_{release} = \varepsilon_{total}$. Stage C adds $\varepsilon_{cert}$.*

The proof (Appendix A) has five composition steps. (i) Forming the cohorts reads only public domains, so it is free. (ii) Within a cohort, the class-balance query and the $|A|$ histograms of a record's own class block form a chain of $q$ Laplace queries of sensitivity 1, which compose sequentially to $\varepsilon_{\mathrm{marg}}$. The two class blocks do not overlap, so they compose in parallel. (iii) Across cohorts, the queries compose in parallel, because each record lies in exactly one cohort. (iv) Within a conditional level, the cells partition $D$ and compose in parallel. A cell's two counts compose sequentially to $\varepsilon_L$, and the published rate is post-processing of those two counts and the public floor. (v) The levels describe the same people, so their budgets add up to $\varepsilon_{\mathrm{cond}}$. The cohort sizes cost $\varepsilon_{\mathrm{count}}$, again in parallel across cohorts. Generation, selection and the sub-bin rule read only $R$, the generated rows and fresh randomness. By post-processing, they add nothing, whatever the pool factor and however many draws are made. Summing the three parts gives $\varepsilon_{\mathrm{count}} + \varepsilon_{\mathrm{marg}} + \varepsilon_{\mathrm{cond}} = \varepsilon_{\mathrm{total}}$.

## 5. Experiments

We evaluate CoRTeC on three public datasets (Table 2). As reference points, we run AIM and MST (the authors' reference implementations), DP-2Stage, the DP-LLMTGen procedure and Tab-PE, both on our split and under each paper's own protocol. Utility is the measure on which a deployment is judged. We report fidelity beside it in every table, because the synthetic rows themselves are also handed out.

### 5.1 Setup

**Table 2:** Datasets. The main split uses seed 42. The replication seeds 43 and 44 each draw a fresh split and a fresh release.

| dataset | records | train / test | numerical / categorical | target, positive rate | role |
|---|---|---|---|---|---|
| UCI Adult [13, 23] | 32,561 | 26,048 / 6,513 | 6 / 8 | income over 50K, 24.1 % | every comparison, the controls, the ladder, the replication |
| NHANES 2017-2018 [9] | 3,749 adults | 2,999 / 750 | 5 / 4 | HbA1c at or above 6.5 %, 14.2 % | auto-configured clinical dataset, the replication |

| dataset | records | train / test | numerical / categorical | target, positive rate | role |
|---|---|---|---|---|---|
| bank marketing [25] | 45,211 | 36,168 / 9,043 | 5 / 8 | term deposit, 11.7 % | third dataset, one split |

**Reference methods.** The reference AIM and MST are the authors' private-pgm implementation (mbi 1.1). Ganev et al. audit MST and AIM as implemented in the dpmm library, with a fixed marginal structure, and find their empirical privacy close to the stated guarantee [16]. We fit both on the training split with the same public bins that CoRTeC declares, the same $\varepsilon$ and a corrected sampler. They run under $(\varepsilon, \delta)$-differential privacy with $\delta = 10^{-9}$, a slightly weaker guarantee than CoRTeC's pure $\varepsilon$. AIM runs under the stable message-passing oracle with a check that the fitted marginals agree with the noisy measurements. MST runs under the library's default oracle, as do the AIM and MST rows at $\varepsilon = 0.5$ and 1 in Appendix E.7. Appendix D gives the settings, explains the choice of oracle and shows how closely the two oracles agree on the seed-42 split. DP-2Stage runs the authors' code (GPT-2, DP-SGD, $\delta = 10^{-5}$) on our split. For the DP fine-tuning rows, we apply DP-LLMTGen's two-stage LoRA [20] procedure and its weighted cross-entropy loss to Qwen3-14B, Qwen2.5-7B and Llama-2 7B on our split ($\delta = 10^{-5}$). The same backbone therefore appears under two privacy designs. Tab-PE runs the authors' code with a population of 1,000 rows (600 for the NHANES draws of 300 rows) and $\delta = 1/(n \ln n)$. The evaluator its authors use, TabICL [19], scores the Tab-PE protocol rows. Two of these controls complete a two-by-two design over the release and the language model. The release without a model is CoRTeC with the maximum-entropy decoder of Section 4.2. The release with a model is CoRTeC with a frozen language model. The model without a release writes rows from the column names alone. This is the baseline Swanberg et al. use [37]. The permuted-target floor has neither.

**CoRTeC configuration.** One set of settings is used for every experiment reported here. The budget is $\varepsilon = 2$, with $\alpha = 0.2$, $\gamma = 0.02$ and $n_{\text{min}} = 150$. Each batch asks for 25 rows, with the counts stated separately for each outcome class. The pool factor is 3 (2 on NHANES), and every CoRTeC configuration except the plain-decoder ablation passes its draws through the selection step with the sub-bin rule. NHANES and bank marketing use the auto-configurator to choose their cohorts and conditional levels, while Adult uses the hand-written rule in its schema. The NHANES and bank marketing configurations were derived during development with the selection noise drawn from a fixed seed. Their choice of columns is therefore not covered by the guarantee, while the released counts, given that choice, stay within the budget of $\varepsilon = 2$. Each configuration was derived once, on the seed-42 training split, and reused on seeds 43 and 44, and its selection budget of 0.1 came on top of the release's $\varepsilon = 2$. The NHANES configuration was derived before 135 records without an HbA1c measurement were excluded from the file (3,107 training rows rather than 2,999). The auto-configured Adult release described below draws that noise afresh, as the released implementation does, and Appendix E.9 reruns NHANES and bank marketing in the same way. Since the auto-configurator is the shipped default, we also ran it on Adult after the main experiments, for consistency rather than as a search over configurations. On each split, the released implementation's Stage A chooses the cohorts and the hierarchy and charges that choice to the same budget. It spends a count share of 0.05, plus a gate charge for the counts that decide which groups clear $n_{\text{min}}$, instead of 0.02 (Appendix D). The rows of this Adult release, labelled "auto-configured", are decoded only with the maximum-entropy decoder and the selection step. Every CoRTeC row is named by its decoder. The plain decoder, the maximum-entropy decoder without that step, is reported as an ablation to show what the step adds. The language-model backbones are Qwen3-8B, Qwen3-14B and Qwen3-32B (4-bit AWQ), all served locally through vLLM. The fourth is Gemini 3.6 Flash. Qwen3-14B is the primary backbone. A Qwen2.5-7B configuration appears only in the backbone ladder, the paper protocols and the other budgets (Appendices E.4, E.6 and E.7).

**The privacy budget.** Recent papers report $\varepsilon$ of 1 or below as their headline, including the two methods that fine-tune a language model [3, 38]. The benchmark of Chen et al. uses 0.2, 1 and 5 [10]. Our headline of $\varepsilon = 2$ comes from the release itself. On the seed-42 Adult release, each count in a conditional cell is noised at scale 10.2. A cell at the floor of 150 records therefore carries its rate with a noise standard deviation of about 0.1. At $\varepsilon = 1$ and 0.5, that error grows to about 0.2 and 0.4. The conditional table then stops being informative near the floor, and a comparison would measure the noise rather than the methods. We therefore chose $\varepsilon = 2$ as the budget at which a cell of floor size still carries its rate with a noise standard deviation of about a tenth of the range. This is a design choice rather than a measured threshold, and we do not claim that a trade-off seen at $\varepsilon = 2$ disappears below it. Appendices E.6 and E.7 report every method at a budget its own paper uses.

**Metrics.** We measure fidelity in four ways against the full private training table over the public bins. One-way and two-way total variation compare single columns and pairs of columns. Conditional total variation compares $P(y \mid \text{group})$ over two families of groups: the cohort bands that the release describes ("seen") and groups built from columns that share nothing with the conditioning ("held-out"). Utility is train-on-synthetic, test-on-real (TSTR) AUC [32]. We train three classifiers (logistic regression, a random forest and gradient boosting), which we call students, on the synthetic table and test them on the real test split. Appendix B gives the definitions. Each paper protocol in Appendix E.6 uses that paper's own metrics and split.

**Floors, ceiling and repetition.** Every table carries three references. The first is a real sample of $n$ training rows, which a synthetic table of the same size is not expected to exceed. The second is the same rows with the target shuffled. It is a floor that carries no information. The third is the ceiling: the students trained on a real sample of 4,000 training rows (the whole split on NHANES, which has 2,999). A metric that cannot separate the first two references cannot detect a lost link to the target. The conditional and utility measures separate them. The marginal measures do not, so no marginal measure is read on its own (Appendix B). Comparing against simple reference points is a principle DPBench recommends [18]. Every cell is the mean over three draws, and Table 3 also gives their sample standard deviation. The central cells are repeated on seeds 43 and 44 with fresh splits and releases. The paired differences across seeds are reported with their sign count and a paired t-test. Results from a single split or draw are labelled as such and read as signals only. With three seeds, a paired test has little power, so we treat a difference as replicated when its sign is the same at every seed.

## 5.2 Without a language model, the release trains models close to a real sample

**Table 3:** Adult at ε = 2 on the seed-42 split. Fidelity is total variation (lower is better), and utility is train-on-synthetic, test-on-real AUC (higher is better). Each cell is the mean ± standard deviation over three draws (five for the reference MST and a single fit for the ceiling). The real-data references are in italics. The Gemini configuration was run at $n$ = 300 only. The maximum-entropy rows pass a pool three times the size of $n$ through Algorithm 2, and the no-selection row is the plain decoder. The one-shot row is the Gemini backbone writing rows from the column names alone, with no release (Appendix D). The auto-configured row decodes the release that the auto-configurator derives on Adult (Section 5.1) in the same way. Its conditional cells are not comparable (marked n/a, see Appendix E.2). Appendix F gives the membership advantage of every configuration.

***n* = 300**

| condition | 1-way TV ↓ | 2-way TV ↓ | cond. seen ↓ | cond. held-out ↓ | TSTR-LR ↑ | TSTR-RF ↑ | TSTR-GBM ↑ |
|---|---|---|---|---|---|---|---|
| *real sample of 4,000 rows (utility ceiling)* | *0.011* | *0.028* | *0.020* | *0.008* | *0.860* | *0.907* | *0.914* |
| *real sample of n rows* | *0.043 ± 0.005* | *0.107 ± 0.004* | *0.062 ± 0.031* | *0.036 ± 0.001* | *0.830 ± 0.013* | *0.870 ± 0.016* | *0.840 ± 0.022* |
| *permuted target (no-information floor)* | *0.041 ± 0.002* | *0.110 ± 0.000* | *0.152 ± 0.045* | *0.102 ± 0.013* | *0.423 ± 0.200* | *0.485 ± 0.127* | *0.482 ± 0.107* |
| CoRTeC, Qwen3-14B | 0.046 ± 0.011 | 0.139 ± 0.010 | 0.027 ± 0.011 | 0.048 ± 0.016 | 0.807 ± 0.026 | 0.862 ± 0.015 | 0.833 ± 0.019 |
| CoRTeC, Gemini 3.6 Flash | 0.027 ± 0.001 | 0.110 ± 0.001 | 0.013 ± 0.008 | 0.042 ± 0.010 | 0.805 ± 0.013 | 0.863 ± 0.008 | 0.838 ± 0.005 |
| CoRTeC, maximum-entropy decoder | 0.033 ± 0.011 | 0.118 ± 0.006 | 0.014 ± 0.006 | 0.048 ± 0.011 | 0.824 ± 0.009 | 0.866 ± 0.004 | 0.862 ± 0.006 |
| CoRTeC, maximum-entropy decoder, auto-configured | 0.019 ± 0.002 | 0.108 ± 0.001 | n/a | n/a | 0.826 ± 0.016 | 0.873 ± 0.009 | 0.860 ± 0.006 |
| CoRTeC, maximum-entropy decoder, no selection | 0.058 ± 0.001 | 0.146 ± 0.002 | 0.082 ± 0.018 | 0.049 ± 0.011 | 0.818 ± 0.020 | 0.856 ± 0.017 | 0.847 ± 0.009 |
| reference AIM | 0.056 ± 0.005 | 0.122 ± 0.001 | 0.055 ± 0.007 | 0.047 ± 0.015 | 0.845 ± 0.004 | 0.869 ± 0.012 | 0.845 ± 0.012 |
| reference MST | 0.048 ± 0.005 | 0.121 ± 0.004 | 0.140 ± 0.021 | 0.043 ± 0.006 | 0.733 ± 0.051 | 0.802 ± 0.021 | 0.748 ± 0.035 |
| DP-2Stage, GPT-2 | 0.280 ± 0.006 | 0.465 ± 0.007 | 0.147 ± 0.005 | 0.141 ± 0.020 | 0.598 ± 0.090 | 0.607 ± 0.155 | 0.601 ± 0.205 |
| DP fine-tuned Qwen3-14B | 0.106 ± 0.004 | 0.190 ± 0.002 | 0.087 ± 0.020 | 0.071 ± 0.008 | 0.809 ± 0.025 | 0.832 ± 0.010 | 0.815 ± 0.007 |
| Tab-PE | 0.203 ± 0.002 | 0.366 ± 0.004 | 0.134 ± 0.002 | 0.049 ± 0.005 | 0.812 ± 0.026 | 0.856 ± 0.013 | 0.839 ± 0.013 |
| one-shot from the column names, Gemini 3.6 Flash (no release) | 0.148 ± 0.002 | 0.307 ± 0.002 | 0.214 ± 0.014 | 0.136 ± 0.007 | 0.795 ± 0.009 | 0.850 ± 0.001 | 0.831 ± 0.012 |

***n* = 1,000**

| condition | 1-way TV ↓ | 2-way TV ↓ | cond. seen ↓ | cond. held-out ↓ | TSTR-LR ↑ | TSTR-RF ↑ | TSTR-GBM ↑ |
|---|---|---|---|---|---|---|---|
| *real sample of n rows* | *0.026 ± 0.004* | *0.061 ± 0.003* | *0.031 ± 0.012* | *0.019 ± 0.002* | *0.855 ± 0.001* | *0.892 ± 0.002* | *0.883 ± 0.004* |
| *permuted target (no-information floor)* | *0.025 ± 0.002* | *0.069 ± 0.003* | *0.146 ± 0.015* | *0.095 ± 0.003* | *0.511 ± 0.119* | *0.508 ± 0.085* | *0.513 ± 0.065* |

| condition | 1-way TV ↓ | 2-way TV ↓ | cond. seen ↓ | cond. held-out ↓ | TSTR-LR ↑ | TSTR-RF ↑ | TSTR-GBM ↑ |
|---|---|---|---|---|---|---|---|
| CoRTeC, Qwen3-14B | 0.036 ± 0.002 | 0.108 ± 0.004 | 0.022 ± 0.002 | 0.031 ± 0.005 | 0.827 ± 0.016 | 0.860 ± 0.007 | 0.859 ± 0.007 |
| CoRTeC, maximum-entropy decoder | 0.027 ± 0.004 | 0.089 ± 0.003 | 0.008 ± 0.002 | 0.037 ± 0.003 | 0.836 ± 0.006 | 0.873 ± 0.004 | 0.874 ± 0.007 |
| CoRTeC, maximum-entropy decoder, auto-configured | 0.016 ± 0.000 | 0.081 ± 0.001 | n/a | n/a | 0.841 ± 0.003 | 0.884 ± 0.002 | 0.884 ± 0.003 |
| CoRTeC, maximum-entropy decoder, no selection | 0.041 ± 0.003 | 0.106 ± 0.005 | 0.034 ± 0.014 | 0.045 ± 0.009 | 0.826 ± 0.008 | 0.870 ± 0.004 | 0.863 ± 0.011 |
| reference AIM | 0.029 ± 0.003 | 0.072 ± 0.003 | 0.028 ± 0.007 | 0.030 ± 0.005 | 0.854 ± 0.002 | 0.888 ± 0.006 | 0.883 ± 0.010 |
| reference MST | 0.031 ± 0.003 | 0.083 ± 0.004 | 0.132 ± 0.017 | 0.044 ± 0.013 | 0.788 ± 0.029 | 0.818 ± 0.017 | 0.804 ± 0.016 |
| DP-2Stage, GPT-2 | 0.277 ± 0.005 | 0.454 ± 0.006 | 0.138 ± 0.020 | 0.133 ± 0.007 | 0.527 ± 0.113 | 0.485 ± 0.038 | 0.484 ± 0.033 |
| DP fine-tuned Qwen3-14B | 0.100 ± 0.003 | 0.172 ± 0.003 | 0.050 ± 0.010 | 0.054 ± 0.008 | 0.809 ± 0.033 | 0.845 ± 0.007 | 0.836 ± 0.012 |
| Tab-PE | 0.196 ± 0.001 | 0.350 ± 0.002 | 0.101 ± 0.017 | 0.031 ± 0.003 | 0.819 ± 0.011 | 0.854 ± 0.011 | 0.843 ± 0.007 |
| one-shot from the column names, Gemini 3.6 Flash (no release) | 0.146 ± 0.002 | 0.305 ± 0.003 | 0.191 ± 0.009 | 0.137 ± 0.005 | 0.807 ± 0.004 | 0.860 ± 0.007 | 0.844 ± 0.012 |

**Decoded without a language model, the auto-configured release is within 0.02 AUC of a real sample.** On Adult, we decode the auto-configured release with the maximum-entropy decoder and the selection step. Over three splits, it trains every student to within 0.02 AUC of a real sample of the same size. On the seeded NHANES and bank marketing configurations, the maximum-entropy decoder was run only without the selection step. There, the gap to a real sample stays within 0.05 AUC: at most 0.048 on NHANES on average over three splits and 0.047 on bank marketing in its one split (Appendices E.2 and E.5). Rerun with freshly derived configurations and with the selection step, the largest gap is 0.031 on NHANES on average over three splits and 0.039 on bank marketing (Appendix E.9). The gain is at $n = 300$. At $n = 1{,}000$, the rerun's largest gap is larger than the seeded rows', 0.031 against 0.027 on NHANES and 0.039 against 0.025 on bank marketing. Table 3 gives the seed-42 split, Figure 3 the means over three seeds and Appendix E.2 every paired test.

On Adult at $n = 1{,}000$, the plain decoder and the Qwen3-14B configuration differ by at most 0.005 in either direction on average over the three seeds. This holds for single columns, pairs, the seen conditionals and every student. None of these differences reaches $p < 0.1$. On the held-out conditionals, Qwen3-14B is better by 0.009 ($p = 0.07$). At $n = 300$, the plain decoder is worse on the seen conditionals (0 of 3 seeds in its favour, $p = 0.04$) and better on the gradient-boosting student (3 of 3, $p = 0.01$). On NHANES, the plain decoder fits pairs more closely than Qwen3-14B at every seed and scores higher on every student at $n = 1{,}000$. On bank marketing, it matches the real sample on single columns in its one split (Appendix E.5).

Every language-model configuration of CoRTeC passes its pool through Algorithm 2. Doing the same with the maximum-entropy decoder's pool removes the plain decoder's shortfall on the seen conditionals, and the result holds on all three splits. With the selection step, the maximum-entropy decoder has lower error than the plain decoder on every fidelity measure except the held-out conditionals at $n = 300$ (0 of 3 seeds against, $p \leq 0.03$). Its error is below that of a real sample of the same size on the seen conditionals at both sizes and on single columns at $n = 300$. The release that the auto-configurator derives on Adult improves further when it is decoded in the same way. On the seed-42 split, its single-column error is 0.019 at $n = 300$ and 0.016 at $n = 1{,}000$, against a real sample's 0.043 and 0.026. Its students score 0.826, 0.873 and 0.860 at $n = 300$, against the real sample's 0.830, 0.870 and 0.840.

**Beside the recent DP generators.** On our split, the students of the maximum-entropy configurations score at or above those of DP-2Stage, the DP fine-tuned Qwen3-14B and Tab-PE at both sizes (Table 3). Under each of those papers' own protocols, CoRTeC's utility is within one percentage point of the method's published result or above it with a language model, and within 1.4 points with the plain decoder (Appendices E.6 and E.7). On these benchmarks, the cost of fixing the workload in advance falls mainly on pairs of columns (Section 5.4).

## 5.3 What each part contributes

**Most of the gain comes from the selection step, not from the model.** On single columns, the auto-configured release improves on the plain decoder by 0.038 at $n$ = 300 and by 0.025 at $n$ = 1,000. Of that, the selection step accounts for 0.023 and 0.014, and the auto-configured workload for 0.015 and 0.011, each with 0 of 3 seeds against ($p \le 0.03$). On the students, each step adds 0.003 to 0.009, significant only for the selection step on the random forest at $n$ = 1,000 ($p = 0.01$).

Compared with the plain decoder, the model's overall effect is neutral on Adult. It is negative on NHANES pairs at every seed ($p \le 0.02$ at both sizes) and on bank marketing at both sizes (Appendix E.5). The language-model configuration's largest consistent gain over the plain decoder is on the seen conditionals on Adult at $n$ = 300 (3 of 3 seeds, $p = 0.04$), and that gain comes from the selection step, not from the model. Against CoRTeC with the maximum-entropy decoder, which also passes its pool through Algorithm 2, Qwen3-14B has higher error on single columns, pairs and the seen conditionals at every seed ($p \le 0.04$ at both sizes) and is level or lower on every student (Appendix E.2). Pair by pair, the model's error is below the plain decoder's on pairs it recognises as logically linked. On Adult, examples are education with education number and marital status with relationship. On bank marketing, one is days since the previous campaign with the previous outcome. Its error is above the plain decoder's on pairs it believes are related when the data says they are not, such as education with contact and loan with contact on bank marketing (Appendix E.3). The model's errors therefore follow its expectations about the columns. Without the release, the Gemini 3.6 Flash backbone writing rows from the column names alone is worse than the shuffled floor on the seen conditionals. About 70 % of its rows are verbatim copies of the first 25 rows of the public file (Appendix F.3), yet its students stay within 0.025 AUC of every language-model configuration of CoRTeC. On a public benchmark, fidelity therefore separates the configurations while utility barely does. Swanberg et al. likewise find such rows inaccurate on pairs [37]. With the release, the same backbone copies none of those rows.

**Take away the column names and the model's advantage goes with them.** The anonymised schema renames every column, category and label to a neutral token. The release and the prompt are mapped in the same way, and the draws are mapped back before scoring. The model therefore sees the same numbers with no meaning attached (Appendix C). Table 4 (Appendix C) gives the results on Adult at $n$ = 300. On Gemini 3.6 Flash, the pair error rises to the plain decoder's level, and the students fall below the plain decoder's. Only the advantages that come from the selection step survive. The single-column error stays at 0.030 and the seen-conditional error at 0.014, against the plain decoder's 0.058 and 0.082, because selection matches the pool to the released counts whatever the names. On Qwen3-14B, six of the seven measures are worse without names, and only the linear student is slightly higher (0.811 against 0.807). On Qwen3-32B, over two draws, all seven are worse. DP-LLMTGen's authors report the same for fine-tuning [38].

## 5.4 Replication, and the cost of a fixed workload

**The results replicate over three splits, and the cost of the design falls on pairs of columns.** Figure 3 shows the CoRTeC configurations on Adult at seeds 42, 43 and 44 beside the band that a real sample of the same size spans. The release spends its budget on single columns within cohorts and on the conditional rates a deployment names. An adaptive workload such as AIM's spends it on the pairs that matter most in the data. The reference AIM fits pairs more closely than CoRTeC with Qwen3-14B at every seed, on Adult and NHANES and at both sizes. Against the maximum-entropy decoder (without the selection step on NHANES), it does so on NHANES at both sizes and on Adult at $n$ = 1,000. On bank marketing, the plain decoder fits pairs at least as closely as AIM (0.066 against 0.068 and 0.043 against 0.049). On utility, the release decoded without a language model reaches the reference AIM in five of the six dataset-size settings. We count AIM's utility as reached on a student when the mean is higher than AIM's, or within 0.01 AUC with no consistent sign across the seeds. The exception is NHANES at $n$ = 300, where AIM's linear student scores higher than the plain decoder's at every seed (by 0.035, $p = 0.06$). This count takes the auto-configured release on Adult and the plain decoder on NHANES and bank marketing. With the fixed-rule release on Adult, decoded with or without the selection step, AIM's linear student scores higher at both sizes, so the count is three of the six.

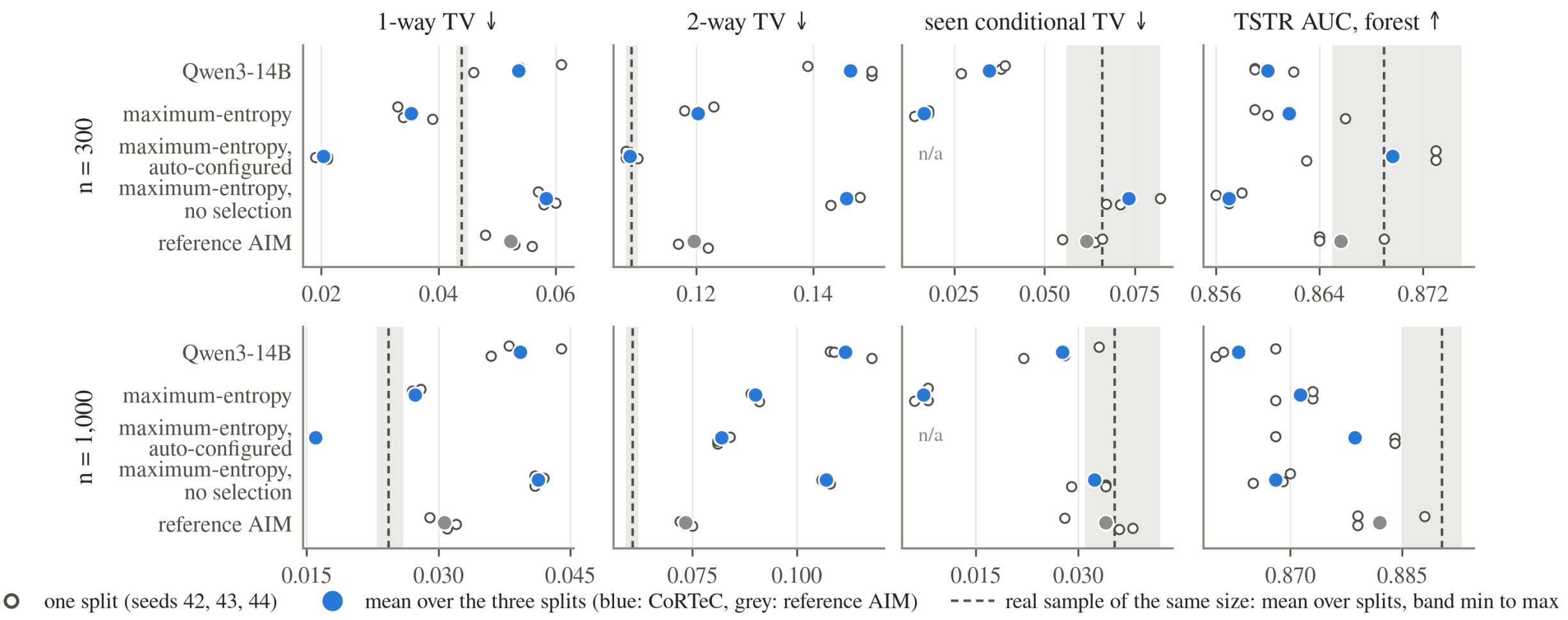

**Figure 3:** Replication on Adult over seeds 42, 43 and 44, each with a fresh split and a fresh release. An open circle is one split's mean over three draws. A filled dot is the mean over the three splits. The upper four rows are CoRTeC, named by decoder, and "no selection" is the plain decoder. The dashed line and the grey band show a real sample of the same size: its mean and its range over the splits. The auto-configured release's conditional cells are not comparable (n/a). Appendix E.2 gives every seed and every paired test.

On Adult, the fixed-rule release carries the seen conditionals more closely than the reference AIM at both sizes ($p < 0.01$ at $n = 300$). AIM's students score higher than those of CoRTeC with Qwen3-14B at $n = 1{,}000$ (p = 0.07 to 0.08), and higher than the maximum-entropy decoder's on two of the three students (p = 0.02 and 0.06). With the auto-configured release, that utility gap closes at both sizes (p = 0.28 to 0.86), and its gradient-boosting student is ahead of AIM at $n = 300$ (3 of 3 seeds, p = 0.03). The two Adult releases differ in cohort size. On the seed-42 split, the fixed rule forms twelve cohorts of 380 to 6,041 records, whereas the auto-configurator forms three cohorts of 4,434 to 13,345. A Laplace histogram has the same absolute noise whatever the count. Larger cohorts give smaller relative errors, while the fixed rule's smaller cohorts are the groups whose conditional rates a deployment names. The conditional claim belongs to the fixed rule, and the utility and single-column claims to the auto-configurator.

### 5.5 The privacy architecture and the cost of a draw

**No private row reaches the model, and nothing is copied into a draw.** Five backbones decoded the same Adult release with the same prompts and settings (Appendix E.4). Within Qwen3, single-column fidelity improves with model size and levels off at 14 billion parameters, and the served Gemini model is best. Utility does not change with the backbone. The same Qwen3-14B fine-tuned with DP-SGD at the same budget is further from the private table on every fidelity measure at both sizes (Table 3). Where the budget is spent matters more than which backbone is used. Membership inference (Appendix F) runs four attacks [11, 33, 34] on 3,000 member candidates from the training split and 3,000 held-out candidates (750 of each on NHANES), and each attack is validated on a positive control. The strongest advantage is at most 0.035 on every Adult configuration and 0.026 to 0.060 across the NHANES configurations of Appendix F, against the 0.762 that the budget permits. The distance-to-closest-record test finds no row in its configurations that copies a training row and no configuration that is closer to the training rows than to a held-out set (Appendix F). The served Gemini backbone can recite rows 2 to 25 of the public Adult file, but none of those rows appears in any CoRTeC draw (Appendix F.3). The release itself takes seconds, and the maximum-entropy decoder runs on a CPU. Appendix G gives a summary in percent.

## 6. Limitations

The NHANES and bank marketing configurations were chosen with selection noise drawn from a fixed seed during development, so their choice of columns is not covered by the guarantee (Section 5.1). We reran those two datasets through the released implementation, with freshly derived configurations and the selection step, together with the membership attacks (Appendix E.9). The plain-decoder and language-model rows of the main tables were not rerun. The auto-configurator also reads the number of training records and the set of observed category levels without noise, so its guarantee treats both as public. Our research pipeline likewise took each categorical column's levels from the training split rather than from a declared schema. On every split we ran, these were exactly the levels of the full public file, so each release is the one a declared schema would give. As a mechanism, though, this is pure ε-DP only if the observed levels are treated as public. Adult's Holand-Netherlands is held by one of 32,561 rows, and a split without that row would release no such level. The released implementation takes the levels from the declared schema. The fixed workload costs accuracy on pairs of columns, which the reference AIM fits more closely on Adult and NHANES (Section 5.4). AIM's students also score higher on NHANES at $n = 300$ and, against the fixed rule, on Adult at $n = 1{,}000$. AIM also fits the seen conditionals more closely on NHANES at $n = 300$ and, against the plain decoder, on bank marketing. Our draws are at most 1,000 rows, so we have not measured utility at larger sizes. The gap to a real sample is largest on the medical dataset, at up to 0.048 AUC on average over three splits (0.068 on the seed-42 split alone). In the rerun, one split reaches 0.073 at $n = 1{,}000$ (Appendix E.9). The language model's contribution depends on the column names, so an institution whose schema the model does not know should expect the maximum-entropy decoder's result rather than the Gemini configuration's. Several results come from a single split: the backbone ladder, the anonymised and one-shot controls, the plain decoder at $\varepsilon = 0.5$ and 1, the bank dataset and the DP-LLMTGen and DP-2Stage protocol rows. The 8B configuration has two draws on Adult at $n = 1{,}000$, the anonymised 32B configuration has two draws, and Gemini ran at $n = 300$ only. The auto-configured Adult release was decoded only with the maximum-entropy decoder, so we have not measured what a language model would add to it. On the seeded NHANES and bank marketing configurations, the selection step was not run, so those rows show the plain decoder. The Qwen3 backbones were not given the memorisation test. The reference AIM runs on CoRTeC's public bins, a conservative version of the PrivTree-binned AIM that Tab-PE's authors run [39]. We make no claim about high-order correlations, the axis on which Tab-PE's stress test separates methods [39]. Our rerun of the DP-LLMTGen procedure understates that method (Appendix E.6), so its published row is the one to use as the reference. Every real dataset here is a public benchmark. The anonymised control removes what the model knows about the schema but not what it may know about the marginals. Draws that fell short of the requested size are scored as delivered (Appendix D), which costs CoRTeC sampling error that the references do not pay.

## 7. Conclusion

For a regulated institution, the product is the release itself. The budget is spent once, on a file of counts with a workload fixed in advance, and anyone can inspect and decode that file at no further cost. Decoded without a language model, the release trains models within 0.02 AUC of a real sample of the same size on Adult and, rerun with freshly derived configurations, within 0.04 on average on NHANES and bank marketing. With the fixed cohort rule and the selection step, it carries the conditional rates it names more closely than such a sample does. The main argument for the design is its privacy architecture: the model never sees a row, every draw is post-processing and the backbone can change without touching the guarantee. The measured membership advantage of the CoRTeC draws, 0.01 to 0.05, is a small fraction of the 0.762 the budget permits. CoRTeC is meant to sit beside the marginal methods, not to replace them. They suit a deployment that needs pairs of columns, and CoRTeC one that needs named conditional rates, a file it can inspect and decoding with any generator. For a deployment, we recommend decoding with the maximum-entropy decoder and the selection step first, and adding a language model only where the model knows the schema.

## Ethics Statement

This work uses public benchmark datasets and a national health survey released for research. No private institutional data was used. The membership attacks were run on synthetic data we generated. The memorisation tests were run against a served commercial model, using the public Adult file, which that model can partly reproduce. The method is meant to reduce the privacy risk of sharing tabular data. Its guarantee covers the release and its draws, not the model's pretraining corpus, and we state its limits where they arise.

## Reproducibility Statement

The repository (Calyie/cortec) holds the Apache-2.0 reference implementation with its tests, the research pipeline that produced every CoRTeC configuration, the paper and the frozen auto-configurations. Released statistics cannot be regenerated bit for bit, because their noise is drawn fresh: from a cryptographically secure source in the research pipeline and from operating-system entropy in the released implementation. We therefore store every release, and every draw reads its release from the stored file. The reference-AIM wrapper, the maximum-entropy decoder with and without the selection step, the one-shot control, the anonymisation of releases, the scorers and the driver of the rerun in Appendix E.9 are all in the repository. The released implementation decodes a release with no language model through `cortec decode` (Private-PGM and the selection step). The synthetic draws, the per-draw score files, the membership-attack files and the stored releases of every configuration are deposited with the Zenodo record of this paper (10.5281/zenodo.22945315). Every number in this paper can be traced to a stored result file.

## AI-Use Statement

Language models appear in this work as the objects of study: they are the language-model decoders named in the tables. Language-model assistants were also used to draft and edit the text and the code. Every experiment was designed, run and checked by the author, and every number was produced by the described pipeline from stored result files.

## Compute and Cost

The Qwen3 backbones and the DP fine-tuning baselines ran on rented NVIDIA A40 GPUs, using vLLM [45] for generation and Opacus [41] for DP-SGD. The reference AIM and MST fits and the maximum-entropy decoder ran on a CPU, and the Gemini configurations ran through Google's interface. The named and anonymised Gemini 3.6 Flash pairs cost 9.37 dollars in total, the memorisation test 0.33 dollars and the one-shot control 2.52 dollars (156 calls). On Adult, a 300-row CoRTeC draw from its pool of 900 rows took 39 to 43 calls of 25 rows on Gemini and Qwen3-14B, and on NHANES, from a pool of 600, 25 or 26. On Gemini, it costs 1.56 dollars at list price, and on Qwen3-14B it takes about 50 minutes on one rented A40.

## Appendix A. Privacy Analysis

*Proof of Theorem 1.* Let $D$ and $D'$ be neighbours that differ in one record $x$.

(i) *Forming the cohorts is free.* $\Pi$ is a function of public attribute domains, so $\Pi(x)$ is computed from $x$'s own attributes without looking at any other record. The partition does not depend on the data, so releasing it costs nothing.

(ii) *Within a cohort, sequential composition.* Fix a cohort $P$. Each released histogram is a counting query of L1 sensitivity 1, because $x$ falls in exactly one bin of it. Removing $x$ changes every histogram that covers it by 1 in L1 norm, which is why the queries on $P$ compose sequentially. The Laplace mechanism at scale $1/\varepsilon_q$ makes each query $\varepsilon_q$-differentially private. Under the pooled block, the $q$ queries on $P$ compose sequentially to $q\ \varepsilon_q = \varepsilon_{\text{marg}}$. Under the class-conditional blocks, $P$ is split by $y$, so $x$ lies in exactly one of $P \cap \{y = 0\}$ and $P \cap \{y = 1\}$. It is touched by the class-balance query and by the $|A|$ histograms of its own block, but by nothing in the other block. That is again a chain of $q$ queries costing $q\ \varepsilon_q$. The two blocks do not overlap and compose in parallel. The pooled histogram is computed from released quantities only.

(iii) *Across cohorts, parallel composition.* The cohorts do not overlap. They are defined by a public rule or, under auto-configuration, by the released output of the selection step, which Theorem 1 charges as $\varepsilon_{\text{sel}}$. The record $x$ lies in exactly one cohort, so it can affect only that cohort's queries. The cost across all cohorts is therefore $\varepsilon_{\text{marg}}$ however many cohorts there are.

(iv) *Within a conditional level, parallel composition.* The cells of a level partition $D$, so the level costs $\varepsilon_L$ however many cells it contains. Within a cell, the positive count and the cell size each have sensitivity 1 under add/remove adjacency. Both are noised with $\text{Lap}(2/\varepsilon_L)$, a scale that depends on nothing private. Since they touch the same records, they compose sequentially to $\varepsilon_L$. The published rate $\tilde{k_c} / \max(\tilde{n_c}, n_{\min})$, clipped to $[0, 1]$, is post-processing of two released counts and the public floor. We do not noise the rate directly with a scale set by the private size $|c|$. Two Laplace densities with different scales have an unbounded likelihood ratio in one tail, so that mechanism would not be pure $\varepsilon$-differentially private.

(v) *Across levels, sequential composition, then the cohort sizes.* The levels describe the same people and are not disjoint, so their budgets add: $|L|\ \varepsilon_L = \varepsilon_{\text{cond}}$. Each cohort size is a counting query of sensitivity 1 noised with $\text{Lap}(1/\varepsilon_{\text{count}})$. The cohorts do not overlap, so all the sizes together cost $\varepsilon_{\text{count}}$. This composes sequentially with each cohort's other queries. Every $n_{\min}$ decision compares a noised count with the public floor. Every cohort and cell of the public domains is queried whether or not it holds a record. Which groups appear in $R$ is therefore post-processing.

*Stage B is post-processing.* The row allocation reads released sizes. The per-bin counts a batch is asked for are apportioned from released histograms. The counts still owed are the difference between released targets and generated rows. The prompt carries $R$ and those counts, and the frozen model reads only the prompt. The acceptance rule reads public domains. The inclusion weights, the raking, the refinement and the sub-bin redraw read $R$, the generated rows and fresh randomness. None of these steps increases the privacy loss at any pool factor or for any number of draws. The maximum-entropy decoder reads only $R$ and is post-processing in the same way.

Adding the parts gives $\varepsilon_{\text{count}} + \varepsilon_{\text{marg}} + \varepsilon_{\text{cond}} = \varepsilon_{\text{total}}$. Under auto-configuration, the noised contingency tables of the candidate columns are counting queries of sensitivity 1 on the same records. They compose sequentially to $\varepsilon_{\text{sel}}$. Everything computed from their noised counts is post-processing. The release that follows spends $\varepsilon_{\text{release}}$, and $\varepsilon_{\text{sel}} + \varepsilon_{\text{release}} = \varepsilon_{\text{total}}$. Stage C's rate queries, one per released cell, are taken over $\max(|c|, n_{\min})$, so each has sensitivity at most $1/n_{\min}$. They act on the disjoint cells of one released level with noise of public scale, and together they cost $\varepsilon_{\text{cert}}$. The ceiling and floor that gate the verdict are scored on a real hold-out sample without noise, so the verdict itself is outside this guarantee. ∎

Floating-point attacks on Laplace [35] and Gaussian [51] samplers apply to every implementation we run, ours included. The release's audit block lists this exposure as an open gap. Ganev et al. audited MST and AIM in another implementation and found them close to their stated guarantee [16].

## Appendix B. Metrics, Floors and Tests

**Fidelity.** The 1-way total variation of a column is half the summed absolute difference between its synthetic and private histograms over the public bins. We average it over columns. The 2-way total variation is the same quantity over the joint histogram of each pair of columns, averaged over pairs. Both are measured against the full private training table. Conditional total variation is the mean absolute difference between $P(y = 1 \mid g)$ in the synthetic table and in the private table over a family of groups $g$. We use two families: the cohort bands of the fixed stratification rule ("seen", the groups that the release describes) and groups built from columns that share nothing with the conditioning ("held-out"). A synthetic table can beat a real sample on the seen conditionals, because the release measures those rates on the full private table, while a sample of *n* rows carries sampling error.

**Utility.** We measure train-on-synthetic, test-on-real AUC with three students: logistic regression (balanced, 2,000 iterations, standardised inputs), a random forest (300 trees, balanced) and gradient boosting (scikit-learn's histogram implementation). Each is trained on the synthetic table and scored on the real test split of the same seed. We report the three students separately, because a result can change sign between a linear and a tree student.

**References in every table.** The real sample is a fresh draw of *n* rows from the training split. It is drawn once per table at that table's *n*, so its value varies between tables by design. The permuted target is the same rows with the target column shuffled. This keeps every marginal and removes every link between the features and the target. The ceiling trains the students on a real sample of 4,000 training rows, or on the whole training split where it holds fewer (NHANES, 2,999 rows). Its fidelity cells are that sample's distance from the full training table: the sampling error at 4,000 rows, or 0 where the sample is the whole split. The real sample and the permuted target separate on every conditional and utility measure but not on the marginal ones. This is why no marginal measure is read on its own.

**Repetition and tests.** Every cell is the mean over three draws from one release, and Table 3 also gives their sample standard deviation. The replication draws a fresh split and a fresh release at seeds 43 and 44. For each pair of conditions, it reports the mean paired difference over the three seeds, the sign count and a two-sided paired t-test (scipy). With three pairs, the test has little power. A difference is therefore read as replicated when its sign is the same at every seed, and its p-value is reported beside it.

**The bin convention.** The evaluator's intervals are closed on the right and the release's on the left. An integer that falls exactly on an edge is therefore counted in different bins by the two. On Adult, this changes 1-way error by 0.003 and changes no ordering.

## Appendix C. Datasets, Preprocessing and the Anonymised Schema

**Adult.** We use the 32,561 rows of the UCI training file and keep "?" as a category. There are 6 numerical columns (age, final weight, education number, capital gain, capital loss, hours per week) and 8 categorical ones. The target is income over 50K. The public bins follow the edges declared in the schema. The cohorts come from the schema's own rule over education and hours per week. The conditional levels are the first and the fourth of the declared hierarchy. The auto-configured Adult release (Section 5.1, Appendix D) takes its cohorts and levels from the auto-configurator instead.

**NHANES 2017-2018.** We join the demographic, glycohaemoglobin, body-measures and blood-pressure files on the respondent identifier and keep adults aged 18 to 80. There are 5 numerical columns (age, BMI, waist, systolic and diastolic blood pressure) and 4 categorical ones (sex, race and ethnicity, education, income bracket). The target is HbA1c at or above 6.5 %. The bin edges are public clinical knowledge: WHO BMI categories and ACC/AHA blood-pressure stages. The release is auto-configured. The auto-configurator's column ranking, run with seeded selection noise during development (Section 5.1), chose three age cohorts and two conditional levels over age, diastolic blood pressure and the race and ethnicity column.

**Bank marketing.** We use the UCI file of 45,211 rows and 13 of its 16 input columns. Five are numerical (age, balance, call duration, campaign contacts, days since the previous campaign) and eight categorical (job, marital status, education, default, housing loan, personal loan, contact type, previous outcome). The day, month and previous-contact-count columns are not used. Call duration is known only once a call has ended, so it partly reveals the outcome. We keep it, as most uses of this file do. The target is a term-deposit subscription, and the split is 80/20 at seed 42. The release is auto-configured in the same way as NHANES.

**The anonymised schema.** A fixed public rule replaces every column name, category label, band label and the dataset description with a neutral token. Numerical columns become n1 to n6 in schema order, categorical columns k1 to k8 and the target y with classes A and B. Each categorical column's sorted categories become k3_v01, k3_v02 and so on. Numerical band labels become n3_b1, n3_b2 and so on. Numeric values, public bounds, bins, the cohort rule, the conditional levels and the held-out families are unchanged. A release of Adult therefore maps one to one onto a release of the anonymised schema. The model sees the same prompt templates with the meaning removed. The draws are mapped back before scoring, so the scorer is the same.

**Table 4:** The anonymised-schema control on Adult at ε = 2 and $n$ = 300, using the same release under neutral names. The anonymised 32B configuration has two draws, and every other row has three. The anonymised draws held fractional values in integer columns, which are floored before scoring as for every draw (Appendix D).

| backbone | names | 1-way TV ↓ | 2-way TV ↓ | cond. seen ↓ | cond. held-out ↓ | TSTR-LR ↑ | TSTR-RF ↑ | TSTR-GBM ↑ |
|---|---|---|---|---|---|---|---|---|
| Gemini 3.6 Flash | real | 0.027 | 0.110 | 0.013 | 0.042 | 0.805 | 0.863 | 0.838 |
| Gemini 3.6 Flash | neutral | 0.030 | 0.148 | 0.014 | 0.044 | 0.745 | 0.840 | 0.798 |
| Qwen3-14B | real | 0.046 | 0.139 | 0.027 | 0.048 | 0.807 | 0.862 | 0.833 |
| Qwen3-14B | neutral | 0.065 | 0.168 | 0.044 | 0.054 | 0.811 | 0.846 | 0.817 |
| Qwen3-32B (AWQ) | real | 0.049 | 0.138 | 0.031 | 0.046 | 0.815 | 0.864 | 0.840 |
| Qwen3-32B (AWQ) | neutral | 0.060 | 0.159 | 0.067 | 0.052 | 0.785 | 0.858 | 0.822 |
| *plain decoder* | *none* | *0.058* | *0.146* | *0.082* | *0.049* | *0.818* | *0.856* | *0.847* |
| *real sample* | | *0.043* | *0.107* | *0.062* | *0.036* | *0.830* | *0.870* | *0.840* |

## Appendix D. Configuration, the Prompt and the Baselines

**CoRTeC settings.** We use ε = 2 (0.5 and 1 where stated), a conditional share α = 0.2, a count share γ = 0.02 and $n_{min}$ = 150. Each call asks for 25 rows, with the counts stated per outcome class. The pool factor is 3 (2 on NHANES), and the selection step uses the sub-bin rule. One cohort worker runs at a time, with seed 42 for the generator and the selection. The Qwen3 backbones ran through vLLM [45] with each model's default sampling. The Gemini backbone ran with reasoning enabled through Google's interface, which has served Gemini 3.6 Flash under the Gemini 3.5 Flash identifier since 8 October 2026. A CoRTeC draw is scored as delivered. Where the pool fell short of the requested size, the draw holds fewer than $n$ rows. On the Qwen3 configurations, this means 276 to 300 rows at $n$ = 300 and 949 to 1,000 at $n$ = 1,000. On the Qwen2.5-7B configuration, the ranges are 257 to 279 and 807 to 864. The Qwen3 ranges include the bank marketing draws (294 to 296 and 973 to 996 rows). On the Gemini configurations, every draw is within 1 % of $n$. The real-sample and permuted references in every table are drawn at the full $n$. Fewer rows add sampling error to a draw's own fidelity measures, so the shortfall works against CoRTeC and against no other method. A call whose output does not parse is retried. The stored manifests record the retry and parse rates for each backbone. The manifests of the ε = 2 Qwen3-14B draws on the three datasets record 2,766 calls, 23 of which returned output that did not parse: 16 on bank marketing, 7 on Adult and none on NHANES. Three bank marketing draws were resumed, and their manifests cover only the final session. Qwen3-8B writes the prompt's range labels into numerical columns when a quota is small. On one NHANES cohort, the quantised 32B model fails about three calls in four, and the run completes when it is resumed.

**Example 1, continued: the cohort in Stage B.** At $n$ = 300 and a pool factor of 3, the cohort is allotted 26 pool rows, in proportion to its released size. The released balance assigns 14 of them to the class above 50K and 12 to the class below. Each class block then fixes how many rows each bin must hold, using the largest-remainder rule. Above 50K, that means three rows aged 40 to 45 and ten husbands. At or below 50K, it means two rows aged 25 to 30 and five people not in a family. The prompt states these counts, and the language model writes rows that carry them. The maximum-entropy decoder needs no prompt. It samples its fitted distribution at the same pool size. The selection step then matches every cohort's pool to the released counts, keeps about nine rows of this cohort in the final 300 and refines them by swapping rows. Finally, it redraws each numerical value inside its released bin. The rows that leave therefore carry the release's structure down to the level of a bin and nothing of the model's below it.

**The prompt.** The experiments used the research pipeline's prompt, which is in the repository with the pipeline. Its system message is generic: it asks for realistic rows that match the column schema, as CSV with one header row and nothing else. The cohort prompt carries the instruction. It gives the dataset's one-line description, the cohort's released size and class balance, the released numerical histograms and categorical shares and the conditional outcome table. In a cohort with class blocks, the histograms and shares are given once for each outcome class. The prompt calls the conditional table the ground truth for this dataset, says that the data may differ from general expectations and asks the model to follow these numbers and not any outside assumption. It then gives the exact counts the batch must carry for each bin, category and outcome, followed by the column order. It ends by asking the model not to use general knowledge about the outcome, to use only the numbers given and to output only CSV with a header row. The reference implementation's template carries the same released numbers in other words, and its system message and template are fixed and fingerprinted. There, the system message asks the model to reproduce the distributions and relationships the statistics describe "even if this contradicts your expectations about how these variables usually relate". The full texts are in the pipeline's source, in the reference implementation's template file and in the deployment guide.

**The integer rule.** Whether a numerical column holds integers is decided from the public bin edges and bounds, which must all be whole numbers, together with the majority of the pool's values. A rule that looks at every value instead lets a single model-written "45.0" or "37.5" turn the column into real values. The sub-bin redraw then produces fractional ages, hours and capital values, which an evaluator comparing exact values treats as far from the data. The anonymised Qwen3-14B draws of Table 4 were selected under that rule and score 0.117 on single columns before flooring and 0.065 after. Every CoRTeC draw in this paper is scored after its integer columns are floored. The released implementation carries the majority rule with a regression test. The Tab-PE and DP-2Stage draws are scored as their methods deliver them, with fractional values in integer columns. Flooring those columns would lower their 1-way and 2-way distances by at most 0.011 and 0.017 on the draws we measured, so no ordering depends on it. Two anonymised Qwen3-32B draws have 25 and 24 empty hours-per-week values, which the scorer drops from that column's histogram.

**Reference AIM and MST.** We use private-pgm (mbi 1.1) at the authors' defaults: an 80 MB maximum model size, a 2-way workload, 1,000 mirror-descent iterations, the exponential mechanism for AIM's selection, $\delta = 10^{-9}$, with the corrected sampler for both. During development, we also ran the smartnoise-synth implementations of AIM and MST at their library defaults. They carry the sampler defect described below, and their students score 0.15 to 0.17 AUC below the reference implementation's on Adult at $n = 300$. No number from them appears in this paper. Every reference AIM and MST number here comes from the authors' private-pgm.

The wrapper is invoked as `pgm_baselines.py --train <split> --dataset <name> --epsilon 2 --n-out <n> --draws 3 --seed 42 --methods aim mst --max-model-size 80 --degree 2 --max-iters 1000 --outdir <folder>`. For the reference AIM rows of Table 3 and of the replication, `PGM_ORACLE=stable` and `PGM_CONSISTENCY_GATE=0.08` are set in the environment. Each seed has three draws under the stable oracle, so the three seeds are like for like. The reference MST rows and the $\varepsilon = 0.5$ and 1 rows of Appendix E.7 are five draws (`--draws 5`) from the wrapper's runs under the library's default oracle. This is why the caption of Table 3 counts five draws for MST. The wrapper is in the repository with the maximum-entropy decoder, the scorers and the anonymisation map.

Under mbi's default message-passing oracle, the fit raised "probabilities contain NaN" inside mirror descent. This happened on the seed-43 and seed-44 splits of both datasets and on seeds 45 to 47, 53, 54 and 60 to 63 in a sweep. The library's fallback, a halved step, returned near-uniform models. On seed 43 (NHANES, $n = 300$), such a model gave a 1-way total variation of 0.182, students at 0.50 and a diabetes rate of 43 % against 14 %. None of those fits is reported. With the log-space oracle (`marginal_oracles.message_passing_stable`), the same seeds converge at the first attempt. On the seed-42 Adult split, the stable refit matches the default-oracle fit to within 0.002 on single columns and pairs (0.029 and 0.072 against 0.031 and 0.074 at $n = 1{,}000$). It matches it to within 0.001 AUC on the students (0.854, 0.888 and 0.883 against 0.853, 0.887 and 0.883). At $n = 300$ the two differ by up to 0.022, on the seen conditionals (0.055 against 0.076) and on the gradient-boosting student (0.845 against 0.824). The consistency check is the mean total variation between the fitted marginals and the noisy measurements they were fitted to. It reads released values only. The library's threshold is 0.15. We set it to 0.08. The first-attempt fits on every seed and size measure between 0.005 and 0.036, so a threshold of about twice the largest accepted value rejects a fallback fit and accepts every legitimate one. Every reported fit passed at its first attempt.

The corrected sampler fixes a defect in mbi 1.1's `synthetic_data`. That function reads each child column's conditional table in the model's attribute order, not in the order in which the parents were sampled. Where the two orders differ, the joint structure of some columns is scrambled while every 1-way marginal survives. The wrapper samples each column from its conditional with the axes aligned. The sampled conditionals then match the fitted model's to within 0.01.

**The plain decoder (the maximum-entropy decoder without selection).** Private-PGM's estimator is given exactly the released cohort sizes, class balances, histograms and conditional table of CoRTeC's release as its measurements, with no further noise and no further query. It is then sampled at the requested $n$.

**The maximum-entropy decoder with the selection step.** On each of the three splits, the maximum-entropy decoder of that split's release is sampled at three times $n$, giving three pools per size. Each pool is passed through Algorithm 2 with the sub-bin rule, using the same seeds as the language-model configurations. The integer columns are then floored. The selection fell one to three rows short of $n$ in six of the eighteen draws. This is post-processing of the release with no privacy cost, and it runs on a CPU.

**The auto-configured Adult release.** On each split, we ran the released implementation's Stage A ( `release_statistics` with the auto-configurator). On the seed-42 and seed-44 splits, it chose three age cohorts and a hierarchy of relationship, then relationship by marital status. On the seed-43 split, it chose education-number cohorts and a hierarchy of relationship, then relationship by native country. The budget went a share of 0.05 to the hierarchy selection and 0.05 to the published counts, plus a gate charge of up to another 0.05 for the noised counts that decide which groups clear $n_{\text{min}}$ (part of $\varepsilon_{\text{count}}$ in Theorem 1). A fifth of the remainder went to the conditional cells and the rest to the marginals. The splits used 104 to 108 queries, and each accounted ε = 2.000 and sealed the ledger. The release was decoded by `cortec decode` (the maximum-entropy fit, a pool of three times $n$ and the selection step) with seeds 42 to 44. The fit's consistency with the release was 0.004. No language model was run on it.

**The one-shot control.** This uses the research pipeline's header-only prompt: the dataset's one-line description, the column names in order, the two labels of the target and a request for 25 realistic rows per call. It uses no release and no private data. The model is Gemini 3.6 Flash with reasoning at its default, as in the Gemini configurations. A 300-row draw takes twelve calls and a 1,000-row draw forty. Integer columns are floored, and the draws are scored as delivered. The draws' positive rate is 36 to 38 % against the data's 24 %. About 70 % of their rows are verbatim copies of the public file's first 25 rows (Appendix F.3).

**DP-2Stage.** We ran the authors' released code: GPT-2, full fine-tuning, DP-Adam, clipping 1, $\delta = 10^{-5}$, loss weight 0.65, five pseudo-data epochs, ten private epochs and temperature 0.7 with rejection sampling. It ran on our seed-42 split at ε = 2. Under the authors' protocol, it also ran on their own split at ε = 1.

**DP fine-tuning of the same backbones.** We applied DP-LLMTGen's procedure to Qwen3-14B, Qwen2.5-7B and Llama-2 7B. Stage 1 trains for five epochs on random tables drawn from the public names, categories and ranges at a learning rate of $10^{-4}$. Stage 2 trains for two epochs of DP-SGD (Opacus [41]) on the private rows, with LoRA [20] of rank 8 on the attention projections (10.5 million trainable parameters on the 14B). It uses a learning rate of $5 \times 10^{-4}$, batch size 32, clipping 1, the weighted cross-entropy with loss weight 0.65 and $\delta = 10^{-5}$. It spent ε = 1.994 of 2. Sampling at temperature 0.7 gave valid rows on 99.7 % of attempts. We omit the original's numerical-understanding loss because its coefficients are not given.

**Tab-PE.** We used the authors' DPSDA code and their Adult example: a mutation rate of 0.5 decaying polynomially to 0.01 with Tab-PE's decay factor of 0.2, the composite population with sampling then ranking, the L2 nearest-neighbour histogram, $\delta = 1/(n \ln n)$, 29 to 30 iterations and a noise multiplier of 11.3 at ε = 2 on our split. Populations were 1,000 rows, except 600 for the NHANES draws of 300 rows. Under the authors' protocol, they were 2,000 rows on their 70/15/15 split with 30 iterations.

## Appendix E. Full Tables

**E.1 NHANES at ε = 2, seed 42, three draws.** Lower is better in the first four columns and higher in the last three. The configuration is the seeded one (Section 5.1), and Appendix E.9 reruns it.

| condition, $n$ = 300 | 1-way TV | 2-way TV | cond. seen | cond. held-out | TSTR-LR | TSTR-RF | TSTR-GBM |
|---|---|---|---|---|---|---|---|
| *all 2,999 training rows (utility ceiling)* | *0.000* | *0.000* | *0.000* | *0.000* | *0.797* | *0.763* | *0.747* |
| *real sample of the same size* | *0.034* | *0.081* | *0.015* | *0.036* | *0.773* | *0.728* | *0.716* |
| *permuted target* | *0.043* | *0.093* | *0.047* | *0.060* | *0.493* | *0.514* | *0.515* |
| CoRTeC, Qwen3-14B | 0.029 | 0.158 | 0.036 | 0.075 | 0.678 | 0.657 | 0.645 |
| CoRTeC, Qwen3-32B (AWQ) | 0.047 | 0.157 | 0.036 | 0.046 | 0.681 | 0.677 | 0.657 |
| CoRTeC, Qwen3-8B | 0.042 | 0.181 | 0.017 | 0.050 | 0.651 | 0.646 | 0.655 |
| CoRTeC, maximum-entropy decoder, no selection | 0.042 | 0.117 | 0.025 | 0.041 | 0.705 | 0.694 | 0.691 |
| reference AIM | 0.042 | 0.100 | 0.017 | 0.044 | 0.747 | 0.721 | 0.710 |
| reference MST (5 draws) | 0.045 | 0.110 | 0.026 | 0.060 | 0.709 | 0.668 | 0.642 |
| DP-2Stage, GPT-2 | 0.175 | 0.295 | 0.117 | 0.115 | 0.527 | 0.537 | 0.553 |

| condition, $n$ = 300 | 1-way TV | 2-way TV | cond. seen | cond. held-out | TSTR-LR | TSTR-RF | TSTR-GBM |
|---|---|---|---|---|---|---|---|
| DP fine-tuned Qwen3-14B | 0.116 | 0.205 | 0.074 | 0.077 | 0.575 | 0.544 | 0.527 |
| Tab-PE (600-row population) | 0.064 | 0.151 | 0.030 | 0.053 | 0.739 | 0.729 | 0.724 |

| condition, $n$ = 1,000 | 1-way TV | 2-way TV | cond. seen | cond. held-out | TSTR-LR | TSTR-RF | TSTR-GBM |
|---|---|---|---|---|---|---|---|
| *real sample of the same size* | *0.016* | *0.039* | *0.008* | *0.016* | *0.783* | *0.752* | *0.744* |
| *permuted target* | *0.017* | *0.044* | *0.039* | *0.047* | *0.516* | *0.484* | *0.470* |
| CoRTeC, Qwen3-14B | 0.027 | 0.150 | 0.031 | 0.069 | 0.696 | 0.648 | 0.639 |
| CoRTeC, Qwen3-32B (AWQ) | 0.036 | 0.123 | 0.023 | 0.031 | 0.748 | 0.727 | 0.701 |
| CoRTeC, Qwen3-8B | 0.034 | 0.156 | 0.023 | 0.040 | 0.748 | 0.688 | 0.671 |
| CoRTeC, maximum-entropy decoder, no selection | 0.031 | 0.088 | 0.022 | 0.029 | 0.758 | 0.720 | 0.686 |
| reference AIM | 0.029 | 0.070 | 0.027 | 0.032 | 0.752 | 0.714 | 0.660 |
| reference MST (5 draws) | 0.028 | 0.075 | 0.016 | 0.048 | 0.727 | 0.692 | 0.679 |
| DP-2Stage, GPT-2 | 0.169 | 0.284 | 0.120 | 0.110 | 0.538 | 0.520 | 0.512 |
| DP fine-tuned Qwen3-14B | 0.117 | 0.195 | 0.089 | 0.078 | 0.571 | 0.516 | 0.501 |
| Tab-PE (1,000-row population) | 0.060 | 0.132 | 0.027 | 0.041 | 0.743 | 0.721 | 0.716 |

**E.2 Replication: means over seeds with their standard deviations, and the paired differences.** The NHANES rows use the seeded configuration (Section 5.1). The real-sample row is each split's own sample of $n$ training rows, averaged over the three splits. It is the band in Figure 3. Each seed's value is the mean of three draws. The p-value is from a two-sided paired t-test over the three seeds. The sign count gives how many seeds have a positive and a negative difference. The maximum-entropy rows are the fixed-rule release decoded with the maximum-entropy decoder and passed through the selection step on each split. The no-selection rows skip that step, and the difference rows call them the plain decoder. The auto-configured rows are the release the auto-configurator derives on Adult, decoded in the same way (Section 5.1). Their conditional cells are not comparable and are left out. The evaluator's seen bands are the fixed rule's cohorts, which that release does not condition on. Two of the three held-out families (relationship, and native country on the seed-43 split) are columns it conditions on. One held-out family shares no column with its conditioning: the income rate by workclass and race. On that family, its conditional total variation, averaged over the three splits with three draws each, is 0.043 against the reference AIM's 0.061 at $n$ = 300 and 0.020 against 0.036 at $n$ = 1,000. No seed favours AIM at either size. The plain decoder reads 0.061 at $n$ = 300 and 0.043 at $n$ = 1,000 on the same family.

| Adult, $n$ = 300 | 1-way TV | 2-way TV | cond. seen | cond. held-out | TSTR-LR | TSTR-RF | TSTR-GBM |
|---|---|---|---|---|---|---|---|
| *real sample of the same size* | *0.044 ± 0.001* | *0.109 ± 0.001* | *0.066 ± 0.014* | *0.043 ± 0.007* | *0.834 ± 0.004* | *0.869 ± 0.005* | *0.841 ± 0.003* |
| CoRTeC, Qwen3-14B | 0.054 ± 0.008 | 0.146 ± 0.006 | 0.035 ± 0.007 | 0.048 ± 0.002 | 0.815 ± 0.009 | 0.860 ± 0.002 | 0.835 ± 0.002 |
| CoRTeC, maximum-entropy decoder, no selection | 0.058 ± 0.002 | 0.145 ± 0.003 | 0.073 ± 0.008 | 0.049 ± 0.003 | 0.818 ± 0.000 | 0.857 ± 0.001 | 0.847 ± 0.002 |
| CoRTeC, maximum-entropy decoder | 0.035 ± 0.003 | 0.120 ± 0.002 | 0.017 ± 0.002 | 0.043 ± 0.004 | 0.822 ± 0.002 | 0.861 ± 0.004 | 0.853 ± 0.008 |
| reference AIM | 0.052 ± 0.004 | 0.120 ± 0.003 | 0.061 ± 0.006 | 0.048 ± 0.001 | 0.832 ± 0.011 | 0.866 ± 0.003 | 0.839 ± 0.006 |
| CoRTeC, maximum-entropy decoder, auto-configured | 0.020 ± 0.001 | 0.109 ± 0.001 | n/a | n/a | 0.830 ± 0.004 | 0.870 ± 0.006 | 0.859 ± 0.002 |
| Qwen3-14B minus plain decoder | −0.004 [1+/2−] 0.47 | +0.001 [2+/1−] 0.88 | −0.038 [0+/3−] 0.04 | −0.001 [1+/2−] 0.44 | −0.003 [1+/2−] 0.63 | +0.003 [3+/0−] 0.12 | −0.012 [0+/3−] 0.01 |
| Qwen3-14B minus AIM | +0.002 [2+/1−] 0.80 | +0.027 [3+/0−] 0.03 | −0.027 [0+/3−] <0.01 | −0.000 [2+/1−] 0.86 | −0.017 [1+/2−] 0.28 | −0.006 [0+/3−] 0.02 | −0.004 [1+/2−] 0.42 |
| Qwen3-14B minus maximum-entropy | +0.019 [3+/0−] 0.02 | +0.026 [3+/0−] 0.01 | +0.018 [3+/0−] 0.02 | +0.005 [3+/0−] 0.20 | −0.006 [1+/2−] 0.40 | −0.002 [1+/2−] 0.33 | −0.018 [0+/3−] 0.08 |
| plain decoder minus AIM | +0.006 [3+/0−] 0.18 | +0.026 [3+/0−] 0.01 | +0.012 [3+/0−] 0.27 | +0.001 [2+/1−] 0.81 | −0.014 [0+/3−] 0.18 | −0.009 [0+/3−] 0.05 | +0.008 [3+/0−] 0.13 |
| maximum-entropy minus plain decoder | −0.023 [0+/3−] 0.01 | −0.025 [0+/3−] 0.01 | −0.056 [0+/3−] 0.01 | −0.006 [0+/3−] 0.20 | +0.003 [3+/0−] 0.17 | +0.005 [3+/0−] 0.19 | +0.006 [3+/0−] 0.31 |
| maximum-entropy minus AIM | −0.017 [0+/3−] 0.03 | +0.001 [2+/1−] 0.76 | −0.045 [0+/3−] <0.01 | −0.005 [1+/2−] 0.23 | −0.010 [0+/3−] 0.20 | −0.004 [0+/3−] 0.01 | +0.014 [3+/0−] 0.01 |

| Adult, $n$ = 300 | 1-way TV | 2-way TV | cond. seen | cond. held-out | TSTR-LR | TSTR-RF | TSTR-GBM |
|---|---|---|---|---|---|---|---|
| auto-configured minus AIM | −0.032 [0+/3−] 0.01 | −0.011 [0+/3−] 0.05 | n/a | n/a | −0.002 [2+/1−] 0.86 | +0.004 [2+/1−] 0.28 | +0.020 [3+/0−] 0.03 |
| auto-configured minus maximum-entropy | −0.015 [0+/3−] 0.01 | −0.012 [0+/3−] 0.02 | n/a | n/a | +0.009 [3+/0−] 0.13 | +0.008 [3+/0−] 0.08 | +0.006 [2+/1−] 0.28 |

| Adult, $n$ = 1,000 | 1-way TV | 2-way TV | cond. seen | cond. held-out | TSTR-LR | TSTR-RF | TSTR-GBM |
|---|---|---|---|---|---|---|---|
| *real sample of the same size* | *0.024 ± 0.002* | *0.061 ± 0.001* | *0.035 ± 0.006* | *0.025 ± 0.005* | *0.849 ± 0.006* | *0.890 ± 0.005* | *0.885 ± 0.005* |
| CoRTeC, Qwen3-14B | 0.039 ± 0.004 | 0.112 ± 0.005 | 0.027 ± 0.006 | 0.033 ± 0.003 | 0.829 ± 0.002 | 0.863 ± 0.004 | 0.860 ± 0.003 |
| CoRTeC, maximum-entropy decoder, no selection | 0.041 ± 0.001 | 0.107 ± 0.001 | 0.033 ± 0.003 | 0.042 ± 0.003 | 0.825 ± 0.001 | 0.868 ± 0.003 | 0.863 ± 0.003 |
| CoRTeC, maximum-entropy decoder | 0.027 ± 0.001 | 0.090 ± 0.001 | 0.007 ± 0.001 | 0.033 ± 0.004 | 0.832 ± 0.004 | 0.871 ± 0.003 | 0.871 ± 0.002 |
| reference AIM | 0.031 ± 0.001 | 0.074 ± 0.001 | 0.034 ± 0.005 | 0.031 ± 0.001 | 0.847 ± 0.007 | 0.882 ± 0.005 | 0.876 ± 0.006 |
| CoRTeC, maximum-entropy decoder, auto-configured | 0.016 ± 0.000 | 0.082 ± 0.002 | n/a | n/a | 0.840 ± 0.006 | 0.879 ± 0.009 | 0.881 ± 0.006 |
| Qwen3-14B minus plain decoder | −0.002 [1+/2−] 0.52 | +0.005 [3+/0−] 0.27 | −0.005 [1+/2−] 0.38 | −0.009 [0+/3−] 0.07 | +0.004 [3+/0−] 0.13 | −0.005 [0+/3−] 0.20 | −0.002 [0+/3−] 0.25 |
| Qwen3-14B minus AIM | +0.009 [3+/0−] 0.04 | +0.038 [3+/0−] <0.01 | −0.007 [0+/3−] 0.09 | +0.002 [2+/1−] 0.36 | −0.017 [0+/3−] 0.07 | −0.019 [0+/3−] 0.07 | −0.016 [0+/3−] 0.08 |
| Qwen3-14B minus maximum-entropy | +0.012 [3+/0−] 0.04 | +0.021 [3+/0−] 0.02 | +0.020 [3+/0−] 0.04 | +0.000 [1+/2−] 0.97 | −0.003 [0+/3−] 0.36 | −0.008 [0+/3−] 0.09 | −0.011 [0+/3−] 0.07 |
| plain decoder minus AIM | +0.011 [3+/0−] <0.01 | +0.033 [3+/0−] <0.01 | −0.002 [1+/2−] 0.69 | +0.011 [3+/0−] 0.03 | −0.021 [0+/3−] 0.03 | −0.014 [0+/3−] 0.04 | −0.014 [0+/3−] 0.07 |
| maximum-entropy minus plain decoder | −0.014 [0+/3−] <0.01 | −0.017 [0+/3−] <0.01 | −0.025 [0+/3−] <0.01 | −0.009 [0+/3−] 0.03 | +0.007 [3+/0−] 0.05 | +0.003 [3+/0−] 0.01 | +0.009 [3+/0−] 0.07 |
| maximum-entropy minus AIM | −0.003 [0+/3−] 0.05 | +0.017 [3+/0−] <0.01 | −0.027 [0+/3−] 0.02 | +0.002 [2+/1−] 0.60 | −0.014 [0+/3−] 0.02 | −0.011 [0+/3−] 0.06 | −0.005 [0+/3−] 0.16 |
| auto-configured minus AIM | −0.015 [0+/3−] <0.01 | +0.009 [3+/0−] <0.01 | n/a | n/a | −0.007 [1+/2−] 0.36 | −0.003 [1+/2−] 0.56 | +0.004 [3+/0−] 0.35 |
| auto-configured minus maximum-entropy | −0.011 [0+/3−] <0.01 | −0.008 [0+/3−] 0.03 | n/a | n/a | +0.008 [3+/0−] 0.23 | +0.007 [3+/0−] 0.17 | +0.009 [3+/0−] 0.11 |

| NHANES, $n$ = 300 | 1-way TV | 2-way TV | cond. seen | cond. held-out | TSTR-LR | TSTR-RF | TSTR-GBM |
|---|---|---|---|---|---|---|---|
| *real sample of the same size* | *0.037 ± 0.006* | *0.086 ± 0.006* | *0.021 ± 0.006* | *0.032 ± 0.004* | *0.766 ± 0.009* | *0.740 ± 0.010* | *0.721 ± 0.005* |
| CoRTeC, Qwen3-14B | 0.044 ± 0.013 | 0.161 ± 0.011 | 0.021 ± 0.013 | 0.057 ± 0.020 | 0.710 ± 0.033 | 0.689 ± 0.028 | 0.667 ± 0.029 |
| CoRTeC, maximum-entropy decoder, no selection | 0.041 ± 0.001 | 0.117 ± 0.001 | 0.024 ± 0.001 | 0.040 ± 0.002 | 0.717 ± 0.016 | 0.713 ± 0.023 | 0.705 ± 0.015 |
| reference AIM | 0.042 ± 0.001 | 0.103 ± 0.003 | 0.019 ± 0.003 | 0.041 ± 0.007 | 0.752 ± 0.005 | 0.725 ± 0.004 | 0.706 ± 0.005 |
| Qwen3-14B minus plain decoder | +0.003 [2+/1−] 0.78 | +0.044 [3+/0−] 0.02 | −0.002 [1+/2−] 0.77 | +0.017 [2+/1−] 0.26 | −0.007 [1+/2−] 0.75 | −0.024 [1+/2−] 0.21 | −0.037 [0+/3−] 0.17 |
| Qwen3-14B minus AIM | +0.001 [2+/1−] 0.87 | +0.058 [3+/0−] 0.02 | +0.002 [1+/2−] 0.83 | +0.016 [2+/1−] 0.37 | −0.042 [0+/3−] 0.12 | −0.036 [0+/3−] 0.13 | −0.039 [0+/3−] 0.18 |
| plain decoder minus AIM | −0.001 [1+/2−] 0.22 | +0.014 [3+/0−] 0.01 | +0.004 [3+/0−] 0.15 | −0.001 [1+/2−] 0.73 | −0.035 [0+/3−] 0.06 | −0.012 [1+/2−] 0.45 | −0.002 [2+/1−] 0.87 |

| NHANES, $n$ = 1,000 | 1-way TV | 2-way TV | cond. seen | cond. held-out | TSTR-LR | TSTR-RF | TSTR-GBM |
|---|---|---|---|---|---|---|---|
| *real sample of the same size* | *0.018 ± 0.002* | *0.040 ± 0.002* | *0.011 ± 0.002* | *0.016 ± 0.001* | *0.785 ± 0.006* | *0.757 ± 0.005* | *0.749 ± 0.005* |
| CoRTeC, Qwen3-14B | 0.040 ± 0.013 | 0.151 ± 0.008 | 0.022 ± 0.008 | 0.054 ± 0.016 | 0.732 ± 0.031 | 0.686 ± 0.037 | 0.669 ± 0.034 |
| CoRTeC, maximum-entropy decoder, no selection | 0.031 ± 0.002 | 0.087 ± 0.002 | 0.018 ± 0.003 | 0.026 ± 0.003 | 0.769 ± 0.011 | 0.745 ± 0.023 | 0.722 ± 0.032 |
| reference AIM | 0.030 ± 0.002 | 0.071 ± 0.001 | 0.022 ± 0.007 | 0.035 ± 0.016 | 0.734 ± 0.027 | 0.701 ± 0.033 | 0.676 ± 0.022 |
| Qwen3-14B minus plain decoder | +0.009 [2+/1−] 0.34 | +0.064 [3+/0−] <0.01 | +0.004 [3+/0−] 0.25 | +0.028 [3+/0−] 0.09 | −0.037 [0+/3−] 0.10 | −0.059 [0+/3−] 0.06 | −0.053 [0+/3−] 0.08 |

| NHANES, *n* = 1,000 | 1-way TV | 2-way TV | cond. seen | cond. held-out | TSTR-LR | TSTR-RF | TSTR-GBM |
|---|---|---|---|---|---|---|---|
| Qwen3-14B minus AIM | +0.009 [2+/ 1−] 0.33 | +0.080 [3+/ 0−] <0.01 | −0.000 [2+/ 1−] 0.96 | +0.018 [3+/ 0−] 0.21 | −0.002 [2+/ 1−] 0.96 | −0.015 [1+/ 2−] 0.73 | −0.007 [1+/ 2−] 0.81 |
| plain decoder minus AIM | +0.000 [2+/ 1−] 0.85 | +0.016 [3+/ 0−] 0.01 | −0.004 [1+/ 2−] 0.34 | −0.010 [1+/ 2−] 0.47 | +0.035 [3+/ 0−] 0.24 | +0.044 [3+/ 0−] 0.21 | +0.046 [3+/ 0−] 0.07 |

**E.3 Per-pair reading.** For every pair of columns, we compared the pairwise total variation of the model's draws and of the plain decoder's draws against the private table on the seed-42 split. On Adult, the Qwen3 and Gemini models' error is below the plain decoder's on the pairs they recognise as logically linked. For education with education number, the plain decoder reads 0.31 to 0.32 and these models 0.08 to 0.30. For marital status with relationship, they read 0.08 to 0.29 against the plain decoder's 0.41 to 0.43. For relationship with sex, they read 0.03 to 0.11 against 0.22 to 0.23, and for workclass with occupation 0.13 to 0.25 against 0.24 to 0.27. Qwen2.5-7B improves on the two relationship pairs (0.37 to 0.40 and 0.15 to 0.17) but not on workclass with occupation or on education with education number, at either size. On the pairs the release describes only through the cohorts, the model's error is at or above the plain decoder's. On bank marketing, the model repairs days since the previous campaign with the previous outcome. It raises the error on four nearly independent pairs on which the plain decoder matches the real sample: education with contact, contact with the previous outcome, marital status with education and loan with contact. The same pattern appears at both sizes, and on Adult on every Qwen3 backbone and on Gemini.

**E.4 The backbone ladder.** One Adult release and one NHANES release are decoded with the same prompts, settings and evaluator. Only the model changes. Each fidelity cell gives single-column and pairwise total variation, and each utility cell gives the three students.

| backbone | Adult *n* = 300 | utility | Adult *n* = 1,000 | utility | NHANES *n* = 300 | utility | NHANES *n* = 1,000 | utility |
|---|---|---|---|---|---|---|---|---|
| Qwen2.5-7B | 0.110 / 0.220 | 0.81 / 0.85 / 0.83 | 0.089 / 0.178 | 0.81 / 0.86 / 0.85 | 0.045 / 0.160 | 0.64 / 0.64 / 0.61 | 0.039 / 0.145 | 0.66 / 0.64 / 0.62 |
| Qwen3-8B | 0.068 / 0.164 | 0.80 / 0.85 / 0.82 | 0.048 / 0.124 (2 draws) | 0.82 / 0.85 / 0.84 | 0.042 / 0.181 | 0.65 / 0.65 / 0.66 | 0.034 / 0.156 | 0.75 / 0.69 / 0.67 |
| Qwen3-14B | 0.046 / 0.139 | 0.81 / 0.86 / 0.83 | 0.036 / 0.108 | 0.83 / 0.86 / 0.86 | 0.029 / 0.158 | 0.68 / 0.66 / 0.65 | 0.027 / 0.150 | 0.70 / 0.65 / 0.64 |
| Qwen3-32B (AWQ) | 0.049 / 0.138 | 0.82 / 0.86 / 0.84 | 0.035 / 0.103 | 0.82 / 0.86 / 0.85 | 0.047 / 0.157 | 0.68 / 0.68 / 0.66 | 0.036 / 0.123 | 0.75 / 0.73 / 0.70 |
| Gemini 3.6 Flash | 0.027 / 0.110 | 0.81 / 0.86 / 0.84 | not run | | not run | | not run | |
| *maximum-entropy decoder, no selection* | *0.058 / 0.146* | *0.82 / 0.86 / 0.85* | *0.041 / 0.106* | *0.83 / 0.87 / 0.86* | *0.042 / 0.117* | *0.71 / 0.69 / 0.69* | *0.031 / 0.088* | *0.76 / 0.72 / 0.69* |
| *reference AIM* | *0.056 / 0.122* | *0.85 / 0.87 / 0.85* | *0.029 / 0.072* | *0.85 / 0.89 / 0.88* | *0.042 / 0.100* | *0.75 / 0.72 / 0.71* | *0.029 / 0.070* | *0.75 / 0.71 / 0.66* |
| *real sample* | *0.043 / 0.107* | *0.83 / 0.87 / 0.84* | *0.026 / 0.061* | *0.86 / 0.89 / 0.88* | *0.034 / 0.081* | *0.77 / 0.73 / 0.72* | *0.016 / 0.039* | *0.78 / 0.75 / 0.74* |

**E.5 Bank marketing at ε = 2, one split, three draws.** The held-out conditional family is not defined on this schema. The configuration is the seeded one (Section 5.1), and Appendix E.9 reruns it.

| condition | *n* | 1-way TV | 2-way TV | cond. seen | TSTR-LR | TSTR-RF | TSTR-GBM |
|---|---|---|---|---|---|---|---|
| *real sample* | 300 | *0.025* | *0.052* | *0.014* | *0.833* | *0.849* | *0.833* |
| *permuted target* | 300 | *0.027* | *0.056* | *0.096* | *0.415* | *0.467* | *0.463* |
| CoRTeC, Qwen3-14B | 300 | 0.040 | 0.112 | 0.013 | 0.802 | 0.799 | 0.781 |
| CoRTeC, maximum-entropy decoder, no selection | 300 | 0.026 | 0.066 | 0.036 | 0.824 | 0.802 | 0.802 |
| reference AIM | 300 | 0.034 | 0.068 | 0.021 | 0.815 | 0.805 | 0.772 |
| *real sample* | 1,000 | *0.014* | *0.028* | *0.008* | *0.862* | *0.871* | *0.860* |
| *permuted target* | 1,000 | *0.013* | *0.031* | *0.089* | *0.503* | *0.483* | *0.485* |
| CoRTeC, Qwen3-14B | 1,000 | 0.034 | 0.101 | 0.011 | 0.810 | 0.784 | 0.759 |
| CoRTeC, maximum-entropy decoder, no selection | 1,000 | 0.013 | 0.043 | 0.013 | 0.837 | 0.848 | 0.837 |
| reference AIM | 1,000 | 0.024 | 0.049 | 0.006 | 0.839 | 0.824 | 0.805 |
| *real sample of 4,000 rows (utility ceiling)* | | *0.007* | *0.014* | *0.005* | *0.869* | *0.890* | *0.891* |

**E.6 The three papers' own protocols on Adult, ε = 1, one split unless marked.** The published rows are the papers' own. Fidelity is lower-is-better under the DP-LLMTGen and Tab-PE protocols and higher-is-better under the DP-2Stage protocol. Our rows are scored by our implementation of each paper's metrics. Under the DP-LLMTGen protocol, the distance columns are on our scorer's scale, which reads the training set itself at 0.009 against the paper's 0.004. Only the utility columns can therefore be compared across the two sources. Under the Tab-PE protocol, the embedding-space precision and recall depend on the implementation and are left out. Rows marked "three seeds" are means over the three Tab-PE splits.

| protocol and method | fidelity (the paper's measure) | utility (the paper's measure) ↑ |
|---|---|---|
| **DP-LLMTGen protocol** (1-way and 2-way TVD against the test split, XGBoost accuracy and AUC) | | |
| DP-LLMTGen, Llama-2 7B (published) | 0.038, 0.068 | 0.831, 0.879 |
| RAP++ [53] (published) | 0.054, 0.099 | 0.803, 0.807 |
| CoRTeC, Qwen2.5-7B (ours) | 0.060, 0.121 | 0.831, 0.881 |
| CoRTeC, maximum-entropy decoder, no selection (ours) | 0.034, 0.089 | 0.818, 0.888 |
| reference AIM (ours) | 0.018, 0.046 | 0.850, 0.902 |
| reference MST (ours) | 0.018, 0.061 | 0.813, 0.836 |
| DP fine-tuned Llama-2 7B, our run of the procedure | 0.103, 0.177 | 0.795, 0.841 |
| **DP-2Stage protocol** (their split and metric suite: Pair, HIST, CorAcc, F1 and AUC, all in %) | | |
| real data (published) | 97.5, 99.1, 97.3 | 69.9, 91.7 |
| AIM (published) | 77.1, 88.4, 86.4 | 59.6, 86.8 |
| DP-2Stage-O (published) | 72.3, 88.5, 55.4 | 30.4, 61.6 |
| DP-2Stage-O, our run of their code | 63.0, 84.7, 56.4 | 26.2, 59.9 |
| CoRTeC, Qwen2.5-7B (ours) | 70.8, 84.6, 63.6 | 61.8, 87.1 |
| CoRTeC, maximum-entropy decoder, no selection (ours) | 74.3, 87.6, 69.8 | 62.8, 87.1 |
| reference AIM (ours) | 93.1, 97.2, 83.1 | 62.9, 87.7 |
| reference MST (ours) | 92.6, 97.2, 74.2 | 51.5, 84.0 |
| **Tab-PE protocol** (1-WD and 2-WD against the private table, TabICL accuracy and macro F1) | | |
| real data (published UB) | 0.014, 0.027 | 84.41, 75.68 |
| AIM (published) | 0.017, 0.034 | 83.36, 76.10 |
| Tab-PE (published) | 0.049, 0.086 | 82.22, 73.66 |
| Tab-PE, our run of their code, three seeds | 0.044, 0.093 | 82.19, 72.96 |
| CoRTeC, Qwen3-14B, three seeds | 0.040, 0.100 | 81.44, 74.46 |
| CoRTeC, Qwen3-32B (AWQ), three seeds | 0.039, 0.101 | 81.42, 74.98 |
| CoRTeC, Qwen2.5-7B, three seeds | 0.088, 0.176 | 81.75, 74.10 |
| CoRTeC, maximum-entropy decoder, no selection, three seeds | 0.045, 0.109 | 81.97, 76.55 |
| reference AIM, three seeds | 0.022, 0.061 | 81.69, 70.84 |
| reference MST, three seeds | 0.023, 0.062 | 80.86, 64.68 |

The published Tab-PE row's embedding-space precision and recall are 34.3 and 77.3. Measured with a 64-16-64 autoencoder trained for each run, our Tab-PE rerun reads 66.7 and 74.6, and our CoRTeC rows read 63.7 to 72.4 and 71.1 to 83.9 with a language model and 57.5 and 88.1 with the plain decoder. The precision column therefore depends on the implementation and is not compared.

**E.7 Other budgets.** This table gives the reference AIM and MST at ε = 0.5 and 1 under our protocol, with five draws each on Adult and on NHANES. Beside them are CoRTeC with Qwen2.5-7B and the plain decoder at the same budgets, with three draws (the plain decoder's ε = 2 rows repeat Table 3, and the Qwen2.5-7B ε = 2 rows repeat Appendix E.4). The NHANES block adds Tab-PE at ε = 0.5 and 1 and has no CoRTeC rows. The DP-LLMTGen protocol is also shown at ε = 0.5. The ε = 2 rows of Table 3 are the reference. At ε = 0.5 and 1, the plain decoder improves on the 7B configuration on single columns, pairs and every student at both sizes. That configuration's errors therefore come from the backbone and not from the release. The reference AIM fits single columns and pairs more closely than the plain decoder at every budget. At $n$ = 300 and ε = 0.5, the plain decoder's students score higher (0.819, 0.855 and 0.850 against 0.811, 0.814 and 0.787, one split).

| **Adult, *n* = 300** | **ε** | **1-way TV** | **2-way TV** | **cond. seen** | **cond. held-out** | **TSTR-LR** | **TSTR-RF** | **TSTR-GBM** |
|---|---|---|---|---|---|---|---|---|
| reference AIM | 0.5 | 0.050 | 0.126 | 0.061 | 0.084 | 0.811 | 0.814 | 0.787 |
| reference AIM | 1 | 0.054 | 0.125 | 0.078 | 0.051 | 0.823 | 0.859 | 0.830 |
| reference MST | 0.5 | 0.053 | 0.129 | 0.132 | 0.054 | 0.774 | 0.810 | 0.780 |
| reference MST | 1 | 0.051 | 0.127 | 0.144 | 0.042 | 0.741 | 0.798 | 0.754 |
| CoRTeC, Qwen2.5-7B | 0.5 | 0.127 | 0.255 | 0.066 | 0.060 | 0.754 | 0.826 | 0.809 |
| CoRTeC, Qwen2.5-7B | 1 | 0.115 | 0.223 | 0.122 | 0.067 | 0.763 | 0.826 | 0.798 |
| CoRTeC, Qwen2.5-7B | 2 | 0.110 | 0.220 | 0.090 | 0.054 | 0.813 | 0.852 | 0.830 |
| CoRTeC, maximum-entropy decoder, no selection | 0.5 | 0.086 | 0.190 | 0.074 | 0.071 | 0.819 | 0.855 | 0.850 |
| CoRTeC, maximum-entropy decoder, no selection | 1 | 0.066 | 0.157 | 0.069 | 0.060 | 0.821 | 0.859 | 0.857 |
| CoRTeC, maximum-entropy decoder, no selection | 2 | 0.058 | 0.146 | 0.082 | 0.049 | 0.818 | 0.856 | 0.847 |

| **Adult, *n* = 1,000** | **ε** | **1-way TV** | **2-way TV** | **cond. seen** | **cond. held-out** | **TSTR-LR** | **TSTR-RF** | **TSTR-GBM** |
|---|---|---|---|---|---|---|---|---|
| reference AIM | 0.5 | 0.037 | 0.087 | 0.037 | 0.031 | 0.841 | 0.880 | 0.873 |
| reference AIM | 1 | 0.033 | 0.080 | 0.051 | 0.032 | 0.837 | 0.873 | 0.861 |
| reference MST | 0.5 | 0.032 | 0.089 | 0.135 | 0.037 | 0.758 | 0.806 | 0.791 |
| reference MST | 1 | 0.032 | 0.086 | 0.136 | 0.044 | 0.781 | 0.817 | 0.802 |
| CoRTeC, Qwen2.5-7B | 0.5 | 0.107 | 0.208 | 0.052 | 0.046 | 0.808 | 0.863 | 0.848 |
| CoRTeC, Qwen2.5-7B | 1 | 0.095 | 0.183 | 0.078 | 0.044 | 0.807 | 0.853 | 0.840 |
| CoRTeC, Qwen2.5-7B | 2 | 0.089 | 0.178 | 0.061 | 0.039 | 0.808 | 0.855 | 0.849 |
| CoRTeC, maximum-entropy decoder, no selection | 0.5 | 0.078 | 0.164 | 0.034 | 0.062 | 0.817 | 0.864 | 0.854 |
| CoRTeC, maximum-entropy decoder, no selection | 1 | 0.052 | 0.123 | 0.036 | 0.048 | 0.819 | 0.867 | 0.857 |
| CoRTeC, maximum-entropy decoder, no selection | 2 | 0.041 | 0.106 | 0.034 | 0.045 | 0.826 | 0.870 | 0.863 |

| **NHANES, *n* = 300** | **ε** | **1-way TV** | **2-way TV** | **cond. seen** | **cond. held-out** | **TSTR-LR** | **TSTR-RF** | **TSTR-GBM** |
|---|---|---|---|---|---|---|---|---|
| reference AIM | 0.5 | 0.053 | 0.122 | 0.025 | 0.046 | 0.716 | 0.681 | 0.665 |
| reference AIM | 1 | 0.040 | 0.104 | 0.040 | 0.042 | 0.716 | 0.709 | 0.693 |
| reference MST | 0.5 | 0.050 | 0.126 | 0.033 | 0.063 | 0.704 | 0.655 | 0.642 |
| reference MST | 1 | 0.045 | 0.109 | 0.027 | 0.046 | 0.703 | 0.658 | 0.622 |
| Tab-PE (600-row population) | 0.5 | 0.130 | 0.238 | 0.042 | 0.066 | 0.697 | 0.685 | 0.670 |
| Tab-PE (600-row population) | 1 | 0.077 | 0.169 | 0.020 | 0.043 | 0.759 | 0.753 | 0.734 |

| **DP-LLMTGen protocol, ε = 0.5** | **1-way** | **2-way** | **3-way** | **4-way** | **5-way** | **ACC** | **AUC** |
|---|---|---|---|---|---|---|---|
| DP-LLMTGen (published) | 0.058 | 0.095 | 0.126 | 0.154 | 0.183 | 0.833 | 0.887 |
| RAP++ [53] (published) | 0.066 | 0.116 | 0.162 | 0.208 | 0.251 | 0.766 | 0.758 |
| CoRTeC, Qwen2.5-7B (ours) | 0.075 | 0.145 | 0.225 | 0.320 | 0.424 | 0.833 | 0.885 |
| CoRTeC, maximum-entropy decoder, no selection (ours) | 0.049 | 0.112 | 0.193 | 0.293 | 0.409 | 0.819 | 0.882 |
| reference AIM (ours) | 0.021 | 0.057 | 0.106 | 0.173 | 0.259 | 0.821 | 0.845 |
| reference MST (ours) | 0.020 | 0.061 | 0.117 | 0.190 | 0.281 | 0.808 | 0.823 |
| DP fine-tuned Llama-2 7B, our run | 0.103 | 0.181 | 0.253 | 0.329 | 0.412 | 0.772 | 0.815 |
| *training set, our scorer* | *0.009* | *0.025* | *0.057* | *0.112* | *0.190* | *0.877* | *0.931* |

**E.8 Stage C.** We use the rule the implementation ships (`bound_with_controls`, run by `controls/stage_c_on_release.py`). It draws one Laplace estimate per cell of the finest released level at $\varepsilon_{cert} = 1$ and scores the synthetic draws, a real hold-out sample of the same size (the whole hold-out split where that is smaller) and that sample with its target permuted against the same estimates. The tolerance is derived from the release. No verdict is given unless the real sample passes, the permuted sample fails and every cell holds at least 20 synthetic rows. We ran it on two releases from the tables, each on the seed-42 split. On the auto-configured Adult release, with its three 1,000-row draws pooled (3,000 rows), the test discriminates over eight cells. The real sample's worst bound is 0.135 and the permuted sample's 0.298, against a tolerance of 0.148. The synthetic draws are outside the tolerance, with a worst bound of 0.156. Their mean bound over the cells is 0.057, against 0.059 for the real sample. The worst cell is a small one with 29 synthetic rows, whose positive rate is 0.21 against a private estimate of 0.08. On the NHANES rerun release (Appendix E.9), with its three 300-row draws pooled (900 rows), no verdict is given. The real sample's worst bound is 0.237 against a tolerance of 0.092 over eight cells, so the test cannot discriminate. The NHANES hold-out split has 750 rows, and two of the eight cells hold fewer than 20 of them. The synthetic draws' worst bound is 0.226. Both worst bounds come from one cell in which the synthetic draws and the real sample agree (0.34 and 0.35). The private estimate there is low because Stage C takes each rate over $\max(|c|, n_{min})$, which understates the rate of a cell holding fewer than $n_{min}$ records (Appendix A). Without that cell, the real sample's worst bound is still 0.21, so no verdict is given either way. Stage C therefore certified neither configuration. On Adult it found one cell where the decoder's rate is off, and on NHANES the hold-out sample is too small for a verdict at eight cells. Each run cost $\varepsilon_{cert} = 1$ on top of its release's $\varepsilon = 2$.

**E.9 NHANES and bank marketing rerun with freshly derived configurations.** The configurations behind the NHANES and bank marketing rows above were derived with seeded selection noise (Section 5.1). We reran the model-free results for both datasets through the released implementation, the path of the auto-configured Adult release. On each split, Stage A derives the configuration with fresh selection noise, counted inside $\varepsilon = 2$, and sizes its conditional table for the 1,000 records requested. The release is decoded by the maximum-entropy decoder with the selection step, three draws per size, and scored by the evaluator on the same split. Every release accounts $\varepsilon = 2.000$. Its choice of columns is covered by the guarantee if the number of training records and the observed category levels, which the auto-configurator reads without noise, are treated as public (Section 6). The fresh selection chose a different hierarchy on each split. On the seed-42 split it chose BMI and then BMI with systolic blood pressure. On the seed-43 split it chose race and ethnicity and then race with age, and on the seed-44 split BMI and then BMI with age. On bank marketing it chose call duration. The table gives the means over the three NHANES splits and the one bank marketing split, beside the seeded configuration decoded without the selection step and a real sample of the same size. The conditional errors are scored on the seeded configuration's bands, so they are not comparable and are left out. The two fidelity columns do not separate a real sample from the permuted-target floor, so only the utility columns test the link to the target. The real-sample row is drawn at the full $n$ on each split, so it differs slightly from Appendix E.2's, which matches the length of the first language-model draw.

| dataset | $n$ | condition | 1-way TV ↓ | 2-way TV ↓ | TSTR-LR ↑ | TSTR-RF ↑ | TSTR-GBM ↑ |
|---|---|---|---|---|---|---|---|
| NHANES | 300 | rerun, fresh configuration, with selection | 0.020 | 0.097 | 0.769 | 0.737 | 0.720 |
| NHANES | 300 | seeded configuration, no selection | 0.041 | 0.117 | 0.717 | 0.713 | 0.705 |
| NHANES | 300 | *real sample of the same size* | *0.036* | *0.084* | *0.766* | *0.736* | *0.719* |
| NHANES | 300 | *permuted target* | *0.038* | *0.088* | *0.507* | *0.501* | *0.502* |
| NHANES | 1,000 | rerun, fresh configuration, with selection | 0.018 | 0.077 | 0.771 | 0.743 | 0.718 |
| NHANES | 1,000 | seeded configuration, no selection | 0.031 | 0.087 | 0.769 | 0.745 | 0.722 |
| NHANES | 1,000 | *real sample of the same size* | *0.017* | *0.040* | *0.785* | *0.756* | *0.748* |
| NHANES | 1,000 | *permuted target* | *0.017* | *0.043* | *0.524* | *0.490* | *0.477* |
| bank marketing | 300 | rerun, fresh configuration, with selection | 0.004 | 0.041 | 0.829 | 0.835 | 0.805 |
| bank marketing | 300 | seeded configuration, no selection | 0.026 | 0.066 | 0.824 | 0.802 | 0.802 |
| bank marketing | 300 | *real sample of the same size* | *0.025* | *0.052* | *0.833* | *0.849* | *0.833* |
| bank marketing | 300 | *permuted target* | *0.027* | *0.056* | *0.415* | *0.467* | *0.463* |
| bank marketing | 1,000 | rerun, fresh configuration, with selection | 0.002 | 0.032 | 0.845 | 0.838 | 0.821 |
| bank marketing | 1,000 | seeded configuration, no selection | 0.013 | 0.043 | 0.837 | 0.848 | 0.837 |
| bank marketing | 1,000 | *real sample of the same size* | *0.014* | *0.028* | *0.862* | *0.871* | *0.860* |
| bank marketing | 1,000 | *permuted target* | *0.013* | *0.031* | *0.503* | *0.483* | *0.485* |

On average over the three NHANES splits, the rerun's students match or exceed this real sample at $n = 300$, and the largest gap at $n = 1{,}000$ is 0.031, against 0.048 and 0.027 for the seeded configuration. All gaps are taken from unrounded means. On bank marketing, in its one split, the largest gap is 0.028 and 0.039, against 0.047 and 0.025. The gain is therefore at $n = 300$, and at $n = 1{,}000$ the rerun's largest gap is larger than the seeded rows' on both datasets. Single splits vary: on the seed-43 split at $n = 1{,}000$, the gradient-boosting student scores 0.677 against the real sample's 0.750. The rerun changes the configuration, the implementation of Stage A and the selection step at once, so the table does not separate their effects. On Adult, the selection step accounts for most of the gain on single columns (Section 5.3). Membership inference on the three 300-row draws of the seed-42 split, with the four attacks of Appendix F, gives a strongest advantage of 0.044 on NHANES (shadow model) and 0.011 on bank marketing. Every nearest-neighbour interval contains chance, and no draw holds an exact match. The positive control separates on all four attacks, weakly on bank marketing, whose training split has 36,168 rows. The language-model rows were not rerun. The research pipeline's original recipe, run with fresh selection noise on the seed-42 NHANES split, chose diastolic blood pressure, BMI and age and left no cell of its finest table above $n_{min}$, so the pipeline refused the release. We did not draw again with that recipe, since choosing among private derivations would itself use the data. Within the rerun, the seed-42 NHANES split therefore saw two fresh derivations. The refused attempt is itself a private computation on that split, at most $\varepsilon = 2$ (its selection and its noised release, of which only the configuration, a few logged cohort sizes and the refusal were seen), so the rerun's spend on that split, the refused attempt included, is at most $\varepsilon = 4$, and Stage C adds $\varepsilon = 1$ (Appendix E.8), separate from the releases behind the main tables. We moved to the released implementation because the original recipe could not release on a fresh configuration, not because of a measured outcome. That recipe also released the counts at the full $\varepsilon = 2$ with the selection's budget on top, whereas the released implementation counts it inside $\varepsilon = 2$.

## Appendix F. Membership Inference, Distance to Closest Record and Memorisation

**F.1 Membership inference.** We use a population-level membership test in the spirit of Shokri et al. [33] and Stadler et al. [34], rather than Stadler et al.'s per-target game, which retrains the generator with and without each target. There are 3,000 member candidates from the training split and 3,000 held-out candidates, or 750 of each on NHANES, whose held-out split has 750 rows. The adversary sees the three draws of $n = 300$ (900 rows) and must say whether each candidate was in the training set. We run four attacks of the kinds collected in the TAPAS toolbox [21]: nearest-neighbour distance, a discriminative shadow model, exact matching and a per-record likelihood-ratio test in the style of Carlini et al. [11] over sixteen bootstrap density models. The advantage is $2 \cdot |\mathrm{AUC} - 0.5|$, so an attack scoring below chance counts by its distance from chance. The positive control passes 900 random training rows off as synthetic, together with the first 200 member candidates. On Adult, about 104 of the 3,000 member candidates are expected among the 900 random rows, so about 300 members have an exact match, and the control's exact-match advantage is 0.098 as measured. On NHANES it is 0.48. The NHANES rows of this appendix use the seeded configuration (Section 5.1), and Appendix E.9 gives the attacks on the rerun. The other three attacks separate the control on both datasets.

| configuration, ε = 2 | dataset | strongest advantage (attack) | nearest-neighbour AUC (95 %) | exact matches |
|---|---|---|---|---|
| CoRTeC, Qwen3-14B | Adult | 0.022 (shadow model) | 0.491 (0.477 to 0.506) | 0 |
| CoRTeC, Qwen3-32B (AWQ) | Adult | 0.035 (likelihood ratio) | 0.492 (0.478 to 0.507) | 0 |
| CoRTeC, Qwen3-8B | Adult | 0.026 (nearest neighbour) | 0.487 (0.473 to 0.502) | 0 |
| CoRTeC, Gemini 3.6 Flash | Adult | 0.022 (nearest neighbour) | 0.489 (0.475 to 0.503) | 0 |
| CoRTeC, Gemini 3.6 Flash, anonymised | Adult | 0.021 (nearest neighbour) | 0.489 (0.475 to 0.504) | 0 |
| CoRTeC, maximum-entropy decoder, no selection | Adult | 0.025 (nearest neighbour) | 0.488 (0.473 to 0.503) | 0 |
| CoRTeC, maximum-entropy decoder | Adult | 0.024 (likelihood ratio) | 0.493 (0.478 to 0.508) | 0 |
| CoRTeC, maximum-entropy decoder, auto-configured | Adult | 0.025 (likelihood ratio) | 0.488 (0.475 to 0.503) | 0 |
| one-shot from the column names (no release) | Adult | 0.012 (nearest neighbour) | 0.494 (0.480 to 0.510) | 5 |
| reference AIM | Adult | 0.024 (nearest neighbour) | 0.488 (0.474 to 0.503) | 0 |
| reference MST | Adult | 0.032 (nearest neighbour) | 0.484 (0.470 to 0.499) | 0 |
| DP-2Stage, GPT-2 | Adult | 0.021 (shadow model) | 0.493 (0.478 to 0.507) | 0 |
| DP fine-tuned Qwen3-14B | Adult | 0.024 (likelihood ratio) | 0.493 (0.479 to 0.508) | 0 |

| configuration, ε = 2 | dataset | strongest advantage (attack) | nearest-neighbour AUC (95 %) | exact matches |
|---|---|---|---|---|
| Tab-PE | Adult | 0.018 (nearest neighbour) | 0.491 (0.477 to 0.506) | 0 |
| CoRTeC, Qwen3-14B | NHANES | 0.026 (likelihood ratio) | 0.509 (0.478 to 0.537) | 0 |
| CoRTeC, Qwen3-32B (AWQ) | NHANES | 0.049 (likelihood ratio) | 0.516 (0.486 to 0.545) | 0 |
| CoRTeC, maximum-entropy decoder, no selection | NHANES | 0.044 (likelihood ratio) | 0.503 (0.474 to 0.533) | 0 |
| reference AIM | NHANES | 0.048 (shadow model) | 0.506 (0.478 to 0.535) | 0 |
| reference MST | NHANES | 0.039 (shadow model) | 0.504 (0.473 to 0.533) | 0 |
| DP fine-tuned Qwen3-14B | NHANES | 0.060 (likelihood ratio) | 0.473 (0.445 to 0.501) | 0 |
| Tab-PE | NHANES | 0.037 (shadow model) | 0.511 (0.480 to 0.541) | 0 |
| *permitted advantage at ε = 2* | | *0.762* | | |

Every nearest-neighbour interval contains chance except the reference MST's on Adult, which ends just below it. These attacks bound leakage only as far as they are strong. A black-box audit of the kind Annamalai and De Cristofaro give for DP-SGD [2] would return an empirical ε instead of a null result. It is the natural next measurement. On Adult, the nearest-neighbour statistic reads slightly below chance for every method, including every CoRTeC configuration, AIM and MST. The private rows lie slightly farther from the synthetic rows than held-out rows do. This is a property of the split and the statistic, not of any method.

**F.2 Distance to the closest record.** For every synthetic row, we compute the L1 distance to the nearest training row over the 15 Adult columns. Numerical columns are scaled to [0, 1] on the training split. A categorical column counts 1 when it does not match. The reference curve is the real test split measured against the training split. Figure 4 shows the Adult configurations at ε = 2. None of these configurations contains an exact copy of a training row, while the real test split itself contains 0.11 % exact duplicates. The medians are 0.145 for real held-out rows, 0.185 for the Gemini configuration, 0.19 for the reference AIM and 0.34 for the reference MST. They are 0.38 and 0.44 for the Qwen3-14B configuration at the two sizes, 1.18 for the plain decoder, 1.8 for DP-2Stage, 3.6 for Tab-PE and 0.105 for the DP fine-tuned Qwen3-14B. The last of these is the smallest, so we ran a paired check. We compared each row's distance to a random 6,513-row subset of the training split with its distance to the 6,513-row test split. A generator that copies training rows sits closer to the first. One that produces typical rows sits equally close to both. The share of rows closer to the training subset is between 0.47 and 0.54 for every configuration, against the 0.50 expected. The fine-tuned model's small distances therefore reflect typical rows, not copying.

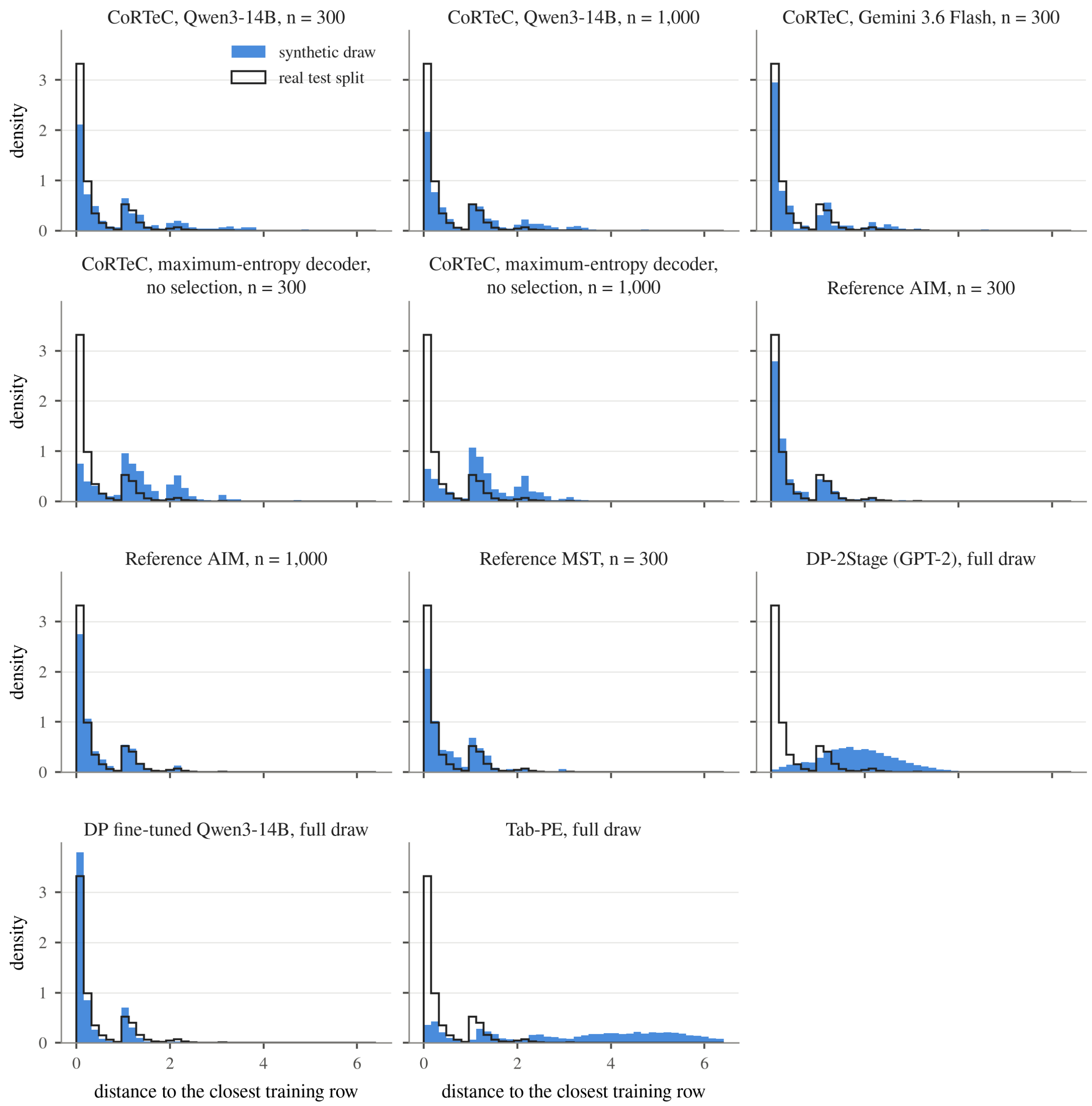

**Figure 4:** Distance to the closest training row on Adult at ε = 2 for each configuration, against the curve of the real test split.

**F.3 Memorisation tests of the served backbone.** We ran the tests of Bordt, Nori and Caruana [4] (package tabmemcheck 0.1.6) on the Adult training file against Gemini 3.6 Flash with reasoning off and temperature 0, in 304 calls. In the header test, the model was given the header and the first characters of row 1. It reproduced data rows 2 to 25 of the file verbatim, including the final-weight values. The opening of the Adult file appears in many tutorials, and the model has memorised it, as Bordt et al. found for GPT-4 [4]. Row completion at random positions with ten real rows as the prefix succeeded 0 times in 25. Completion of the nearly unique final-weight column from the other fourteen succeeded 0 times in 25. The first-token test scored 8 of 25 against 7 of 25 for always choosing the most common token. In feature completion, each categorical column was predicted from the other columns for 25 rows, given five examples. Every column was completed at a rate below, or within the 95 % interval of, what a gradient-boosting classifier trained on the other columns achieves. Education was completed 25 times in 25 against a predictability of 1.00, occupation 7 times against 0.39, workclass 14 against 0.76, relationship 22 against 0.79 and income 21 against 0.87. Every column except workclass was above the majority choice (workclass 0.56 against 0.70). This reflects knowledge of what the columns mean, which the anonymised control removes.

None of the file's first fifty rows appears in any CoRTeC draw on Gemini (named or anonymised) or on Qwen3-14B or 32B at either size. The one-shot control of Table 3, which has no release, shows what that recitation looks like in a draw. Of its rows, 208 to 211 of 300 and 697 to 703 of 1,000 are verbatim copies of rows of the public file (22 to 24 distinct rows, each repeated about nine times at $n$ = 300 and thirty times at $n$ = 1,000). Almost all of them copy the file's first 27 rows, and the rest copy rows 99 and 32,560. Its three $n$ = 300 draws hold five exact matches among the 3,000 membership candidates (three members and two held out). No attack turns them into an advantage (0.012). Against the whole public file of 32,561 rows, no draw of any CoRTeC configuration, of the plain decoder or of the reference AIM holds a verbatim row. Rows that agree on the fourteen columns other than the nearly unique final weight occur in the CoRTeC draws at about the reference AIM's rate (10 to 16 % against 8 to 14 % of a draw at $n$ = 300). That is coincidence on a table with many near-duplicate rows, not copying. The Qwen3 backbones were not given these tests.

## Appendix G. An Indicative Summary in Percent

For a reader who wants one number per axis, we scale fidelity against a real sample of the same size and utility between the no-information floor and the ceiling. Fidelity % is 100 times the mean of min(1, real-sample error / method error) over the four fidelity measures. A value of 100 therefore means at least as close to the private table as a real sample of that size, and 50 means twice as far. Utility % is 100 times the mean of (method AUC − permuted-target AUC) / (ceiling AUC − permuted-target AUC) over the three students, clipped to [0, 100]. Each row averages the configuration's cells at $\varepsilon = 2$, computed from the score files deposited with the draws. Only the cells at $n$ = 300 and 1,000 that the tables above show are included, and the Qwen3-8B row covers Adult only. The cells column counts them. The summary is indicative only and is not a ranking. The tables above are the evidence. Rows are ordered by the number of cells and then by fidelity. Only the rows for the reference AIM, CoRTeC with Qwen3-14B and the three maximum-entropy rows span more than one split. Every other row comes from the seed-42 split alone. The auto-configured release's fidelity mean uses only the two marginal measures, because its conditional cells are not comparable.

| method | cells | fidelity % | utility % |
|---|---|---|---|
| reference AIM | 14 | 78 | 85 |
| CoRTeC, maximum-entropy decoder, no selection | 14 | 72 | 86 |
| CoRTeC, Qwen3-14B | 14 | 68 | 79 |
| CoRTeC, maximum-entropy decoder, auto-configured | 6 | 93 | 93 |
| CoRTeC, maximum-entropy decoder | 6 | 89 | 91 |
| CoRTeC, Qwen3-32B (AWQ) | 4 | 66 | 80 |
| reference MST | 4 | 62 | 71 |
| CoRTeC, Qwen2.5-7B | 4 | 49 | 67 |
| Tab-PE | 4 | 40 | 86 |
| DP fine-tuned Qwen3-14B | 4 | 36 | 49 |
| DP fine-tuned Qwen2.5-7B | 4 | 35 | 47 |
| DP-2Stage, GPT-2 | 4 | 19 | 15 |
| CoRTeC, Qwen3-8B | 2 | 71 | 83 |
| DP fine-tuned Llama-2 7B | 2 | 35 | 64 |
| one-shot from the column names (no release) | 2 | 23 | 85 |
| CoRTeC, Gemini 3.6 Flash | 1 | 96 | 86 |